\documentclass[twocolumn]{article}

\usepackage[
backend=biber,
style=numeric,
sorting=ynt
]{biblatex}
\usepackage{authblk}

\usepackage{multirow}
\usepackage{xcolor}
\usepackage{wasysym}
\usepackage[capitalise, noabbrev]{cleveref}
\usepackage{array}
\usepackage{longtable}
\usepackage{titlesec}
\usepackage{circledsteps}

\usepackage{graphicx}
\usepackage{url}

\newcommand{\ActE}{AE}

\newcommand{\takeaway}[1]{
    \vspace{1em}
    \noindent\fbox{
        \parbox{.96\linewidth}{
            \textbf{Takeaway message:}~#1
        }
    }
}

\newcommand{\x}{$\times$}

\begin{document}

\title{Current Trends in Digital Twin Development, Maintenance, and Operation: An Interview Study

\color{red} \large The final open access SoSyM article available at \url{https://doi.org/10.1007/s10270-024-01167-z}
}

\author[1]{Hossain Muhammad Muctadir\thanks{h.m.muctadir@tue.nl}}
\author[1]{David A. Manrique Negrin\thanks{d.a.manrique.negrin@tue.nl}}
\author[2]{Raghavendran Gunasekaran\thanks{r.gunasekaran@tilburguniversity.edu}}
\author[1, 3]{Loek Cleophas\thanks{l.g.w.a.cleophas@tue.nl}}
\author[1]{Mark van den Brand\thanks{m.g.j.v.d.brand@tue.nl}}
\author[2]{Boudewijn R. Haverkort\thanks{b.r.h.m.haverkort@tilburguniversity.edu}}

\affil[1]{Software Eng.~\& Technology cluster, Department of Mathematics and Computer Science, Eindhoven University of Technology, Eindhoven, The Netherlands}

\affil[2]{Tilburg School of Humanities and Digital Sciences, Tilburg University, Tilburg, The Netherlands}

\affil[3]{Department of Information Science, Stellenbosch University, Stellenbosch, South Africa}
\maketitle

\abstract{
Digital twins (DT) are often defined as a pairing of a physical entity and a corresponding virtual entity (VE), mimicking certain aspects of the former depending on the use-case. In recent years, this concept has facilitated numerous use-cases ranging from design to validation and predictive maintenance of large and small high-tech systems. Various heterogeneous cross-domain models are essential for such systems and model-driven engineering plays a pivotal role in the design, development, and maintenance of these models. We believe models and model-driven engineering play a similarly crucial role in the context of a VE of a DT. Due to the rapidly growing popularity of DTs and their use in diverse domains and use-cases, the methodologies, tools, and practices for designing, developing, and maintaining the corresponding VEs differ vastly. To better understand these differences and similarities, we performed a semi-structured interview research with 19 professionals from industry and academia who are closely associated with different lifecycle stages of digital twins. In this paper, we present our analysis and findings from this study, which is based on seven research questions. In general, we identified an overall lack of uniformity in terms of the understanding of digital twins and used tools, techniques, and methodologies for the development and maintenance of the corresponding VEs. Furthermore, considering that digital twins are software intensive systems, we recognize a significant growth potential for adopting more software engineering practices, processes, and expertise in various stages of a digital twin's lifecycle. 
}

\section{Introduction}
\label{sec:introduction}
Digital Twins (DTs) have captured the interest of industry and academia in recent years, because of their promise to help to better understand, monitor and improve systems. The concept of DTs often encompasses the notion of a real-world entity and a digital counterpart that mimics certain aspects of the former. We believe that both industry and academia are playing a crucial role in further developing the DT concept as well as its design, development, operation, and maintenance. Although growing in popularity, the concepts and practices used around DTs show significantly differences. According to Zhang et al.~\cite{ZHANG2021151} there is no general consensus on the nature of the real-world entity, the required fidelity of the digital counterpart, or the terms used to refer to these entities.

In recent times, an increasing number of domains are developing and/or using DTs to address use-cases spanning over a large spectrum of complexity. A significant number of these DTs are related to complex industrial systems that utilize cutting-edge technologies to execute complex tasks with an extremely high degree of precision and efficiency. Various cross-domain and heterogeneous models plays a central role within these systems and the use of model-driven engineering (MDE) techniques are paramount in the design, development, and maintenance of these models, which is clearly apparent from the current literature~\cite{baker2005, DEARAUJOSILVA2021101021, DAVIES201488, RODRIGUESDASILVA2015139}. We believe that the role of models and MDE is equally crucial in the development of the digital counterpart or virtual entity (VE) of these physical industrial systems in the context of DTs. Our viewpoint is based on three facts: firstly, models are widely recognised as abstractions of real-world concepts, which aligns with the overarching objective of VEs and, therefore, integration of diverse cross-domain models into a VE is logical; secondly, reuse of existing models and related artifacts can significantly improve the development speed of the corresponding VE; finally, the general benefit of MDE, which makes it increasingly popular~\cite{DAVIES201488}, also apply to the development of VEs.

\begin{figure}[ht]
    \centering
    \includegraphics[width=\linewidth]{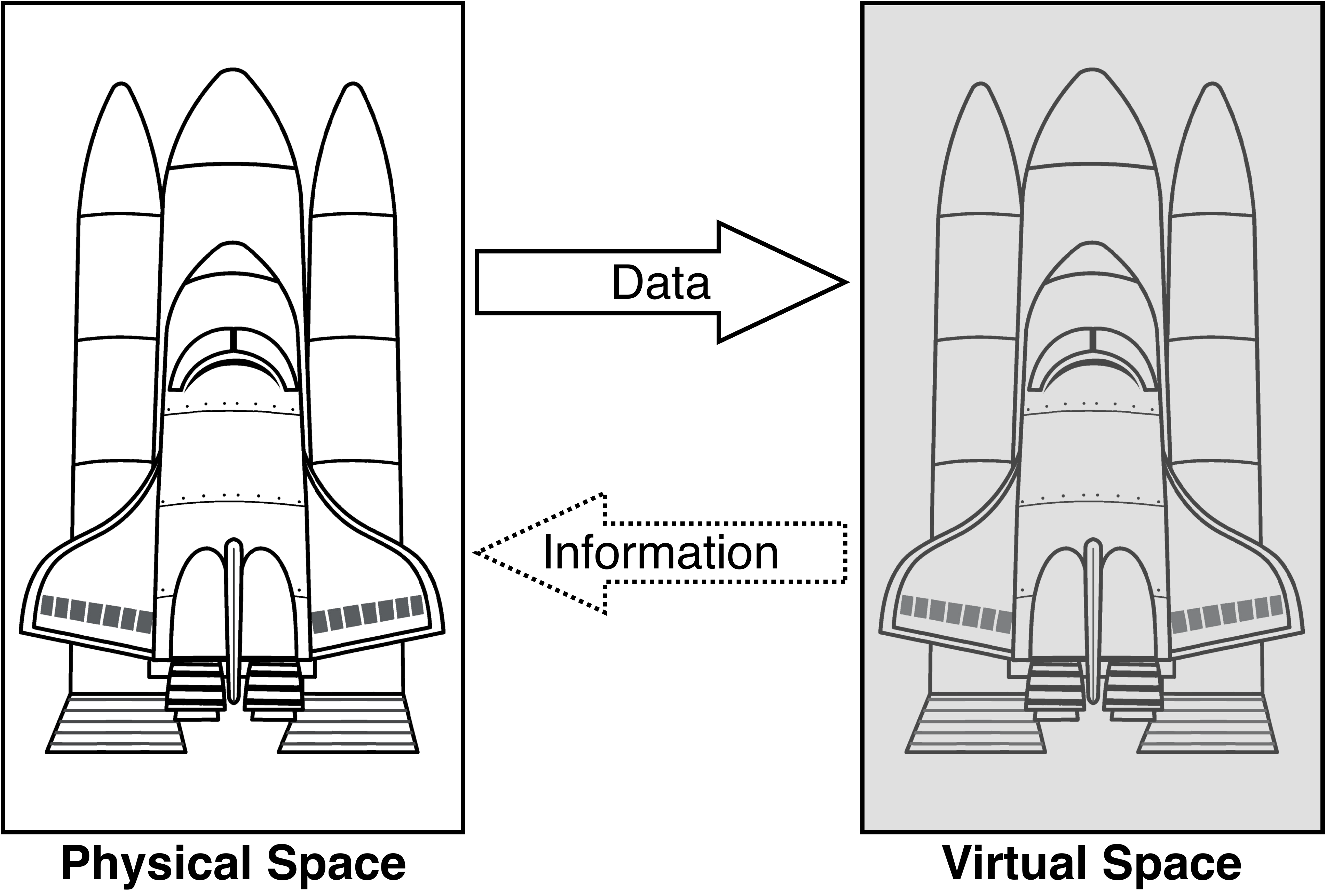}
    \caption{Depiction of the concept of digital twin as introduced by Grieves~\cite{grieves2017digital} where he conceptualized a virtual space or virtual entity (VE) and synchronisation with its physical counterpart. Our research focuses on the VE.}
    \label{fig:grieves_dt}
\end{figure}

As model-based systems, VEs consist of cross-domain and heterogeneous models, which must be (KC1) kept consistent, (KC2) properly orchestrated, and (KC3) validated for correct behavior. Although these activities are identified as key challenges in~\cite{Models_meet_data}, they have not received sufficient attention in the current literature, as far as we are aware. We believe this shortfall can be attributed to the vast diversity of DTs and their rapid large-scale adoption both in industry and academia. In this research, we aimed to better understand the concepts, practices, and methods around the design, development, and maintenance of VEs. To achieve this, we interviewed 19 individuals from industry and academia who are all involved in the design, development, maintenance, and use of DTs of complex industrial systems. The interviews and subsequent analysis provided us with practical insights into diverse understanding of DTs and practices around the three earlier mentioned challenges. Our work was guided by the following seven research questions (RQ):

\begin{itemize}
  \setlength{\itemindent}{2em}
  \setlength{\itemsep}{4pt}  
        \item [RQ 1:] How are digital twins defined in practice?
        \item [RQ 2:] How does reuse of existing (software) artifacts influence the lifecycle of DTs?
        \item [RQ 3:] How is consistency maintained among cross-domain models used in a DT yet developed independently?
        \item [RQ 4:] What technologies and methodologies are used to integrate models in a DT?
        \item [RQ 5:] What practices are used to design and develop the orchestration and data exchange between models in DTs?
        \item [RQ 6:] What techniques and tools are used to validate a DT and its overall dynamic behavior?
        \item [RQ 7:] What properties need to be validated in a DT for its consistent dynamic behavior?
\end{itemize}

\noindent RQs 3--7 are directly related to the three key challenges (i.e., KC1--3) mentioned earlier. In contrast, RQs 1--2 intend to understand the perception of DTs in practice and the role of reusing various artifacts in the creation of VE and its overall impact on DT lifecycle. We believe insights gathered based on these two RQs are valuable to understand the insights from the latter RQs and interesting for not only the DT but also the MDE community at large.

We structure this paper as follows. \cref{sec:Background} describes the background of this research, \cref{sec:motivation_contribution} elaborates further on our motivation for performing this exploratory research and points out the core contributions of this paper. The detailed explanation of the research methodology we followed is explained in \cref{sec:methodology}. In \cref{sec:interviewee_demographics}, we present a short summary of the domains covered by the interviewees and corresponding DT applications. Based on the analysis of the information we gathered during the interviews, in \cref{sec:results} we present our findings related to the above RQs; this section is subdivided into sub-sections, each one devoted to a single RQ. Additional findings that are not strictly related to the RQs but are relevant in the context of this paper are presented in \cref{sec:additional_findings_ext}. \cref{sec:threats} explains the threats to validity of our research and how we attempted to minimize them. In \cref{sec:discussion} we present a discussion correlating findings from individual RQs, recent work that explores various technical aspects of DTs, and our observations. \cref{sec:conclusion} concludes the paper.

\section{Background} %
\label{sec:Background}
The concept of Digital Twins (DTs) was introduced by Grieves~\cite{grieves2017digital} in 2003; Grieves modelled DTs with three dimensions i.e., the physical entity, virtual model and connection, which facilitates the physical–virtual interaction. Since then researchers and practitioners have used DT as an umbrella term to refer to something from a simple simulation to a complex virtual entity closely mimicking a real-world counterpart. For example, Bielefeldt et al.~\cite{bielefeldt2015computationally} focus on ultra-realistic multi-physical computational models in their definition of DT, whereas El Saddik~\cite{el2018digital} defines a DT as a digital replica of a physical entity whether living or non-living. 

\begin{figure}[!ht]
    \centering
    \includegraphics[width=\linewidth]{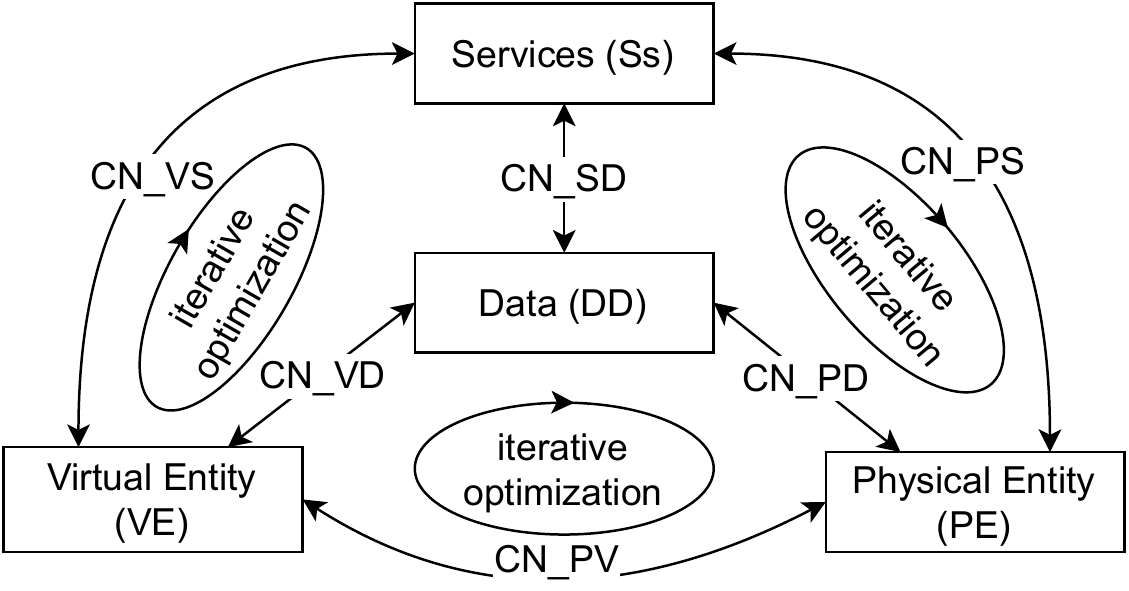}
    \caption{Five dimensional (5D) model of a DT proposed by Tao et al.~\cite{tao2017digital} }
    \label{fig:5d_DT}
\end{figure}

Tao et al.~\cite{tao2017digital} extended the original DT model by Grieves and proposed a five dimensional (5-D) model depicted in Figure~\ref{fig:5d_DT}, i.e.,~$M_{DT}$ = (PE, VE, Ss, DD, CN) where PE refers to the physical (actual) 
real-world entity with various functional subsystems, VE is the corresponding high-fidelity digital model that reproduces certain abilities and properties of the PE, Ss represents the services for PE and VE, DD encapsulates the domain knowledge--data from both PE \& VE and their fusion, and finally, CN is the connection among parts of the DT. 

In order to avoid any ambiguity, in the following sections we use the terms actual entity (\ActE) and virtual entity (VE) to respectively refer to real-world entity, which can be an existing or foreseen engineered or naturally occurring physical system or process, and the corresponding digital counterpart.

Since the inception of DTs, research has focused on understanding the concept, the development of DT's applications, or exploration of different implementation technologies used. An example is the work of Liu et al.~\cite{DT_concept_tech_applications} that analyses the concept of a DT, the technologies used, and DTs' main industrial applications. That research is based on systematic literature review (SLR) analysing over 200 publications.

Tao et al.~\cite{Tao_DT-in-Industry} analyses the state-of-the-art in development and applications of DTs, aiming to outline key technologies enabling DT development, classify current and future applications, and lay out possible gaps and challenges. 
The
 research used an SLR, analyzing 50 papers and eight patents. 

Sharma et al.~\cite{DT_state_art_1}, based on an SLR of over 80 papers, analysed and proposed solutions for the research and implementation gaps of DT technology, such as IoT (internet of things), machine learning and data. This research concluded that regulation and security mechanisms are essential for the proper implementation of DTs due to its cross-domain nature. They also concluded that there are multiple technical and domain specific challenges that require more research to be resolved.

G\"urd\"ur et al.~\cite{interviews_DT_Infrastructure} explores how DTs can help the infrastructure industry. The research methodology used semi-structured expert interviews with non-technical executives from industry in the UK. This approach allowed the researchers to collect their opinions, related to non-technical challenges, on the value of DTs.

Dalibor et al.~\cite{Cross-Domain_SystMapping} conducted an SLR on 356 papers, analysing DTs with a bottom-up approach exploring different implementations to investigate expected DT properties and how DTs are deployed, operated and evaluated. In addition, the authors developed a DT feature model. They explored different implementation techniques, tooling and development processes.

The majority of the research shown above is based on SLRs focused on DT practices and understanding from a high-level systems perspective. The empirical research by G\"urd\"ur et al. also approached DTs from a business and high-level systems perspective.

\section{Motivation and contribution} \label{sec:motivation_contribution}

Although Grieves~\cite{grieves2017digital} introduced the concept of DT more than two decades ago, practitioners are discovering its potential only recently, resulting into an increase of activities and practices around DTs. An increasing number of DTs are being developed and used in domains such as automotive, healthcare, manufacturing, and construction. The application of these DTs extends across a wide spectrum of use-cases, ranging from basic monitoring to highly intricate control systems and predictive maintenance applications. Due to the diversity of use-cases and domains involved, the tools, methods, and practices around the design, development, and maintenance of DTs vary significantly. The central objective of this research is to understand and study this diversity. Our work concentrates on the design, development, and maintenance of the virtual part of a DT, which we refer to as virtual entity (VE) earlier. 

VE, which is often described as digital replication of certain capabilities of a physical entity, is essentially a complex software- and data-intensive system that encompasses various heterogeneous and multi-domain models~\cite{Wright2020, Tao_DT-in-Industry}. To maintain the continued operation of a DT, the corresponding models must be kept consistent with each other, they need to be orchestrated properly, and finally, the dynamic behaviour of both the individual models and the DT as a whole must be validated. Although these tasks are well-known and researched within the broader model-driven engineering (MDE) community, we are unaware of any work focusing on these topics in the context of DTs. This deficiency motivates our work to interview 19 DT practitioners from industry and academia, in order to acquire insight into current challenges, methods and practices related to the three previously mentioned activities (i.e., KC1--3) in the context of DTs. 

Our research was guided by the seven research questions mentioned in \cref{sec:introduction}. Among these research questions, RQs 1--2 delve into the wide-range of understanding of the DT concept and impact of reusing various artifacts on the lifecycle of DTs, respectively. RQ 3 focuses on consistency challenges among heterogeneous and multi-domain models, RQs 4--5 are about the integration and orchestration of these models, and RQs 6--7 concentrate on validation techniques and properties of DTs. Through the interviews and the subsequent analysis of the corresponding transcripts we identified various practices, methods, and tools related to the seven RQs, challenges faced by practitioners, and implemented measures addressing these challenges. Furthermore, we also discussed unresolved challenges, indicating possible research directions, and the similarities and differences between industrial and academic practices.

\section{Research methodology} \label{sec:methodology} %
\begin{figure*}[!ht]
    \centering
    \includegraphics[width=.8\linewidth]{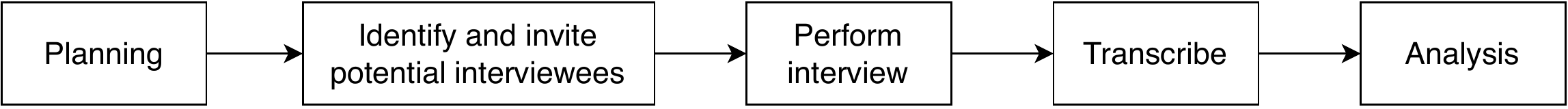}
    \caption{An overview of our major research activities.}
    \label{fig:research_steps}
\end{figure*}

Considering the exploratory nature of our research questions we opted for semi-structured interviews~\cite{magaldi2020semi}. This provided sufficient flexibility for the participants to express themselves while allowing us to collect data on our topics of interest. In this section, we introduce and explain our research methodology, following Strandberg's interview lifecycle~\cite{Strandberg2019EthicalEngineering}, with the steps depicted in Figure~\ref{fig:research_steps}. We expand on each step and explain the related activities in the following subsections.

\subsection{Planning}
This phase entailed the regulatory activities that our universities required for collecting data from human participants, and preparing a questionnaire consistent with the RQs, serving as guideline during the interviews.

\subsubsection{Ethical review and research data management} \label{sec:erb_process}
Ethical review is a process followed by our universities that enables researchers to perform research activities in accordance with accepted ethical standards and existing regulations. This process ensured that measures and infrastructure were in place for maintaining data security and confidentiality as we collected personally identifiable information for (prospective) interviewees and recorded the interviews which potentially contained sensitive information.

\subsubsection{Questionnaire design} \label{sec:question_design}
We prepared a questionnaire to act as a guideline to keep the discussion in our semi-structured interviews focused. %
It was developed in line with the Interview Protocol Refinement (IPR) framework~\cite{castillo2016preparing}, comprising four phases:

\begin{enumerate}
    \item \textbf{Ensure interview questions are aligned with RQs:}
    We took an iterative approach. For the first iteration, we listed our initial RQs and from these derived the initial set of interview questions. We tagged the interview questions with corresponding RQs. This allowed us to identify under-represented research questions and adapt the interview questions accordingly. We repeated this step until the questionnaire stabilized.
    
    \item \textbf{Constructing an inquiry-based conversation:}
    We categorized the interview questions into (1) background, (2) key, and (3) concluding ones. Based on this categorization and suggestions of Hove and Anda~\cite{Hove2005ExperiencesResearch}, we sorted them and rephrased some, enabling an inquiry-based
 conversation. 
    
    \item \textbf{Receiving feedback on the questionnaire:}
    We performed several review rounds among the authors of this paper and a pilot interview with a researcher working in the model-driven software engineering domain to check how well participants did understand the questionnaire. Wherever we identified significant difference between interviewee perception and our intention, we rephrased our questions for better understandability.
    
    \item \textbf{Conduct pilot interview:}
    We performed mock interviews with colleagues, allowing us to pilot our questionnaire, receive feedback, and gather experience as interviewers. This helped to further mature the questionnaire.
\end{enumerate}

\noindent \cref{tab:questions} shows the resulting set of interview questions and their associations with the RQs. All the interview questions except for the first two are connected to one or more RQs. These two questions allowed us to start the conversation, get acquainted with the interviewee, and contributed to a conversation-like interview. Furthermore, with the final question we asked the interviewees' opinion on Tao et al.'s 5-D DT model~\cite{tao2017digital} (see \cref{sec:Background}). While asking this question, we showed an image of the 5-D model and briefly explained it. To avoid influencing the interviewees during the rest of the interview, we intentionally asked this question at the end.

\begin{table*}[!ht]
\caption{Interview questions and their relation to the RQs.}
\label{tab:questions}
\begin{tabular}{|p{.02\linewidth}p{.56\linewidth}|l|l|l|l|l|l|l|l|}
\hline
 &
   \multicolumn{1}{r|}{\textbf{Research questions (RQs) $\rightarrow$}}&
  1 &
  2 &
  3 &
  4 &
  5 &
  6 &
  7 &
  8 \\ \hline
\multicolumn{10}{|l|}{\textbf{Background questions} }   \\ \hline
  1. &
  What is the domain of your company and what kind of services does it provide? &
   &
   &
   &
   &
   &
   &
   &
   \\ \hline
  2. &
  How long have you been working and what is your current role? &
   &
   &
   &
   &
   &
   &
   &
   \\ \hline
\multicolumn{10}{|l|}{\textbf{Key questions}} \\ \hline
  3. &
  What is your understanding of a DT? &
  \x &
   &
   &
   &
   &
   &
   &
   \\ \hline
  4. &
  How are you involved in the development or usage of the DT? &
   &
   &
   &
   &
   &
   &
   &
   \\ \hline
  5. &
  What problems are you solving with your DT? &
  \x &
   &
   &
   &
   &
   &
   &
   \\ \hline
  6. &
  Is your DT for the entire process/system or a specific part? &
  \x &
   &
   &
   &
   &
   &
   &
   \\ \hline
  7. &
  What are the main parts of your DT? Could you shortly describe their role? &
  \x &
   &
   &
  \x &
   &
   &
   &
   \\ \hline
  8. &
  Is there a physical counterpart of your DT? Does it communicate with the digital world? If yes, how? &
  \x &
   &
  \x &
   &
   &
   &
   &
   \\ \hline
  9. &
  Did you build your DT from scratch or you reused some of the things that already existed? What kind of issues did you face  if you reused something or by building from scratch? &
   &
  \x &
   &
   &
   &
   &
   \x &
   \\ \hline
  10. &
  How is the data exchange between the models (and physical system) specified? &
   &
   &
  \x &
  \x &
  \x &
   &
   &
   \\ \hline
  11. &
  How is the sequence of execution of the models in your DT? &
   &
   &
   &
   &
  \x &
   &
   &
   \\ \hline
  12. &
  What issues, if any, did you face with the overall collective execution of models in your DT? &
   &
   &
   &
   &
   &
   &
   \x &
   \\ \hline
  13. &
  Which platforms/tools are you using to develop the digital twin? &
   &
   &
  \x &
  \x &
  \x &
  \x &
   &
   \\ \hline
  14. &
  How often is the DT updated for bug fixes, improvement or similar reasons? &
   &
   &
  \x &
   &
   &
   &
   &
   \\ \hline
  15. &
  After each update did you face integration issues? If yes, what kind? &
   &
   &
  \x &
   &
   &
   &
   \x &
   \\ \hline
  16. &
  How do you ensure various software elements can work together specially if multiple tools were used for development? &
   &
  \x &
  \x &
  \x &
  \x &
  \x &
   &
   \\ \hline
  17. &
  What properties/parameters did you validate to ensure an overall consistent behavior of your DT? &
   &
   &
   &
   &
   &
   &
   \x &
   \\ \hline
  18. &
  What tools and techniques did you use to validate these properties and parameters? &
   &
   &
   &
   &
   &
  \x &
  \x &
   \\ \hline
  19. &
  What do you consider to be the general characteristics of your DT? &
  \x &
   &
   &
   &
   &
   &
   &
   \\ \hline
\multicolumn{10}{|l|}{\textbf{Concluding questions}} \\ \hline
  20. &
  How do you see the DT evolve in the future to solve additional problems? &
   &
   &
   &
   &
   &
   &
   &
  \x{} \\ \hline
  21. &
  What is your opinion on the 5-dimensional DT model from Tao et al.~\cite{tao2017digital}? &
  \x &
   &
   &
   &
   &
   &
   &
   \\ \hline
\end{tabular}
\end{table*}

\subsection{Finding interviewees}\label{sec:find_interviewee}
We aimed to interview practitioners and researchers actively involved in the design, development, maintenance and use of DTs of complex industrial systems. We started with our own network, which consists of a large number of academics and industrial experts working with such systems, as our universities are located within the industrial hub of the Netherlands with many start-ups, high-tech, and manufacturing companies. We initially created a set of potential interviewees that matched our search criteria. Additionally, we verified the DT related involvements of these individuals based on their Google Scholar or LinkedIn profile. Furthermore, as we conducted the interviews, we requested participants to propose potential interviewees from their network, which added two individuals to our list. In the end, we invited 25 persons. We received 22 responses, 20 of them positive; in the end, we conducted 19 interviews, (one respondent stopped responding). Among these interviewees, 10 were from industry and 9 from academia. In \cref{sec:interviewee_demographics} we further discuss the demographics of the interviewees.

We emailed potential interviewees to introduce ourselves and explained the purpose of our research. With the emails, we also included a consent form in which we explained details of our research about data processing and our measures for ensuring data anonymization and privacy. This allowed us to create a level of trust with the interviewees. While we did not a priori  share our questionnaire with the interviewees to avoid prepared and potentially biased answers, we provided them with a description of our topics of interest to allow them  to prepare if they wanted to.

\subsection{Performing interviews} \label{sec:performing_interview}
As we conducted semi-structured interviews, we did not follow any concrete structure; the questionnaire in Table~\ref{tab:questions} acted as a guideline. Therefore, the variation in the quantity of interview questions linked to each RQ had no impact on the quantity of information we were able to gather. Our methodology allowed us to ask more questions if necessary. 

We prepared standard texts that we read to the interviewee at the beginning and end of the interview. These introduced the interviewers, checked whether the interviewee had any questions regarding the consent they provided earlier, and at the end, thanked them for their participation and requested their feedback on the interview. We recorded video and audio of all 19 interviews. These interviews took place between September 2021 and February 2022.

\subsection{Transcription}
The interviews altogether accounted for just over 26 hours of recorded video with audio. To generate word-to-word transcripts of these recordings, we used automated transcript generation followed by manual verification and revision. As most interviews were conducted and recorded using Microsoft Teams, we could use the generated transcript of the corresponding recording. We manually verified and revised each transcript twice to ensure correctness.

\subsection{Data analysis}
\label{sub:data-analysis}
\begin{figure*}[!ht]
    \centering
    \includegraphics[width=.8\linewidth]{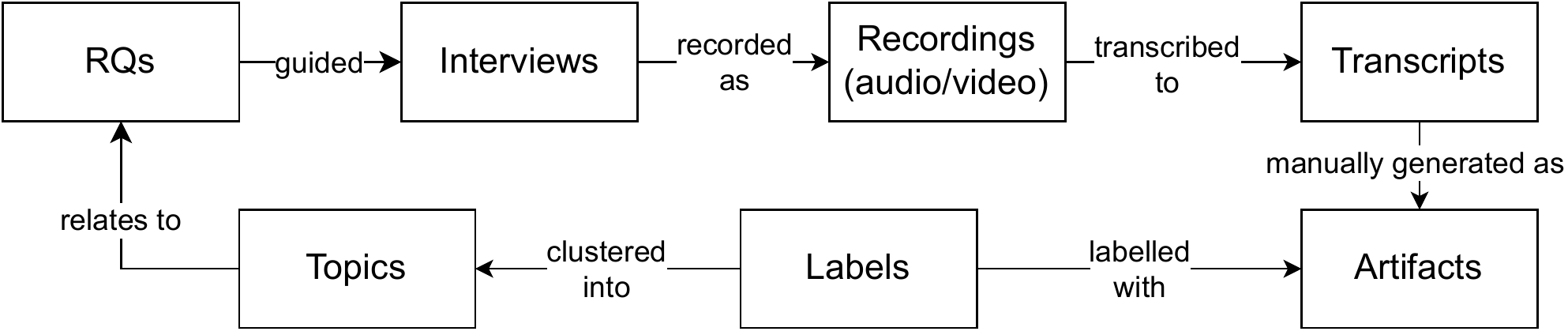}
    \caption{Relationship among various research activities and related outcomes.}
    \label{fig:artifact_relations}
\end{figure*}

We further analysed the transcripts based on the thematic qualitative analysis methodology~\cite{Kiger2020}, the activities related to which are explained in this section. Each of these activities utilize outcomes from one or more of the previous activities. \cref{fig:artifact_relations} depicts the relationships among the activities and the related outcomes.

We used LaMa~\cite{lama2023}, a web-based tool for collaborative labeling and thematic analysis, to collaborate on this analysis. To restrict access and ensure data privacy, we deployed this platform locally.

\subsubsection{Generating and anonymizing artifacts}
The aim of this step was to generate a set of \textit{artifacts} from the transcripts of the interviews. We define an artifact as an independent piece of text that focuses on a specific subject and contains sufficient context information for understanding that subject. To generate them we manually went through each transcript focusing on text spoken by the interviewee and separated text fragments whenever we identified different subjects being discussed. At this stage we only tried to identify changes in subject, not subjects themselves. It was interesting to see that the change of subject occurred not only when new questions were asked but also while discussing one single question. As we generated these text fragments, we kept sufficient context information for them to be understandable. When this was not the case, we added a few keywords as context, marking such an addition with square brackets, e.g., to indicate that the word ``they'' (at that point) refers to a ``[digital entity and its 3D visualization]''. We also generated artifacts by splitting one artifact into two or more during the labeling step, which is explained in the next section. Typically, we split artifacts if we found more than one key message in it. At the end, we had 748 text artifacts of various sizes. Furthermore, we anonymized the transcripts during this manual artifact generation: all personally identifiable information was replaced by unique identifiers that we stored separately for traceability purposes. The anonymization was essential for performing unbiased analysis in later phases of our research.

\subsubsection{Labeling of artifact and topic generation} \label{sec:label_and_topic_generation}
After generating the artifacts 
for all transcripts, we labeled them using LaMa~\cite{lama2023}. We define a label as a short text that sufficiently captures the core message of an artifact. To reduce bias in labeling, each artifact was labeled by two labelers. We resolved conflicting labels by agreeing on one label through discussion.
During the labeling process, the labeler could use an existing label or create a new one. In LaMa, these labels were accompanied by a description explaining how and when a label should be used, which was crucial for reuse of existing labels. Moreover, while labeling we encountered artifacts that lacked sufficient information or context. We labeled these artifacts with two predefined labels: \textit{No value} or \textit{Not understandable}. An example is an artefact discussing an interviewee research policy in relation with their clients. This artefact was classified as \textit{No value}, since it does not discuss any information related to DT development.  

Once we had over 70\% of the artifacts labeled, we started with topic creation in parallel. In this case, we define a {\em topic} as a clustering of labels that can collectively provide a complete message on a specific subject. We chose an iterative approach to create the topics. Based on our initial overview of the existing labels, we created our first set of topics. The topic \textit{Different understandings of DT} is one of the first topics we created. We created this topic because we noticed more than 50 labels related to DT definition. These topics and the list of labels were revisited at regular intervals, resulting in one or more of (1) creation of new topics, (2) redefining a topic at a higher level of abstraction, (3) breaking up a topic into multiple topics, and (4) moving labels from one topic to another. We repeated this until reaching convergence.

\subsubsection{Relating topics to RQs and perform analysis}
In this last step, we focused on answering the research questions. To do that we created a matrix with the final set of topics and our RQs that enabled us to identify the correlations between the two. This matrix allowed us to aggregate and group relevant information from the topics, corresponding labels and artifacts. From this we identified opinions, trends, and practices, which formed the foundation for answering the research questions. To reduce individual opinions and bias, we only considered information groups where three or more interviewees contributed their opinions. However, we occasionally deviated from this while discussing challenges mentioned by the interviewees as we were able to identify limited number of challenges related some of the RQs, which is expected considering the infancy of the DT concept. Excluding these outliers would also mean losing the corresponding valuable insights.

\subsection{Replication package}

The replication package for this interview study is available at~\cite{muctadir_2023_10187933}. Due to the agreement made with our interviewees, we must keep the interview transcripts confidential. Although this significantly constraints on the reproducibility and validation of the findings presented in this paper, we believe that the provided package can still significantly contribute to similar research endeavours. This package includes:
\begin{itemize}
    \item \textbf{Ethical review form:} We used this form for the ethical review process, which is explained in \cref{sec:erb_process}.
    \item \textbf{Interview questions:} This document contains the set of interview questions that guided the semi-structured interviews. These are identical to the ones listed in \cref{tab:questions} and explained in \cref{sec:question_design}.
    \item \textbf{Participation consent form:} We used this form for collecting a written and formal consent from the potential interviewees. To do that we introduced ourselves and explained about this interview study including the purpose of this study, interview procedure, nature of the data we intended to collect, format of the collected data (i.e., audio, video), usage of this data, and confidentiality and privacy measures we put in place to protect this data. 
    \item \textbf{Labels and related topics:} This document contains the \textit{labels} and \textit{topics}, which are explained in \cref{sec:label_and_topic_generation}, we used to annotate the interview data.
\end{itemize}

\section{Demographics of interview participants} \label{sec:interviewee_demographics}

As explained in Section~\ref{sec:methodology}, our aim is to interview individuals who are actively involved in the development, maintenance, and use of DTs. Out of the 19 interviewees, ten were primarily from industry and nine from academia. In this section, we provide an overview of their domains and DTs applications.

We classified the professional domains of the interviewees into six categories. Seven interviewees claimed that their work involves two different domains and one mentioned being involved in three domains. Table~\ref{tab:interviewees_domain} shows the distribution of domains among the 19 interviewees. As visible here there are two dominant categories; the manufacturing and chemical process industry category, and the high-tech products category.

\begin{table}[!ht]
\caption{Professional domains of the interviewees. The number represents the number of DT's in that domain. }
\label{tab:interviewees_domain}
\centering
\begin{tabular}{l|c}
\hline
\multicolumn{1}{c|}{Domain}          & \# DTs \\ \hline
Manufacturing and chemical processing & 10     \\
High-tech products                   & 9      \\
Building and construction             & 3      \\
Transport and logistics               & 3      \\
Information systems                  & 2      \\
Healthcare                            & 1      \\ \hline
\end{tabular}
\end{table}

All the interviewees mentioned that they have multiple applications for their DTs. We therefore analysed the correlation between the domain and the DT's application. For the analysis, we classified the applications in eight categories, as follows (in alphabetical order).

\begin{itemize}
    \item \textbf{Analysis} or improvement of the operation of a product or process. Examples provided by interviewees are bug detection, optimization, system behavior analysis and simulation for decision making or process configuration. 
    \item \textbf{Control} of a process or a product. Such applications aim to take corrective actions from a monitored state towards a desired one.
    \item \textbf{Demonstration} of alternative solutions or configurations for a physical system. This tends to use visual tools, such as 3D modeling tools.%
    \item \textbf{Design and development} of a product (hardware or software) or process. Examples presented by interviewees are prototyping, use of simulation for design and improvements. 
    \item \textbf{Monitoring} a property within a process or a product, aiming to compare the current state against a planned one.
    \item \textbf{Predictive maintenance} concerns the supervision and prediction of an equipment of product condition. The aim of such a DT application is to determine the state of health of the supervised entity and anticipate its maintenance. 
    \item \textbf{Testing} products or processes; examples shared by interviewees are virtual commissioning, verification and validation or experimentation.    
    \item \textbf{Training} operators or users of a specific machine or product.
     
\end{itemize}

\begin{figure*}[h!]
    \centering
    \includegraphics[width=0.8\textwidth]{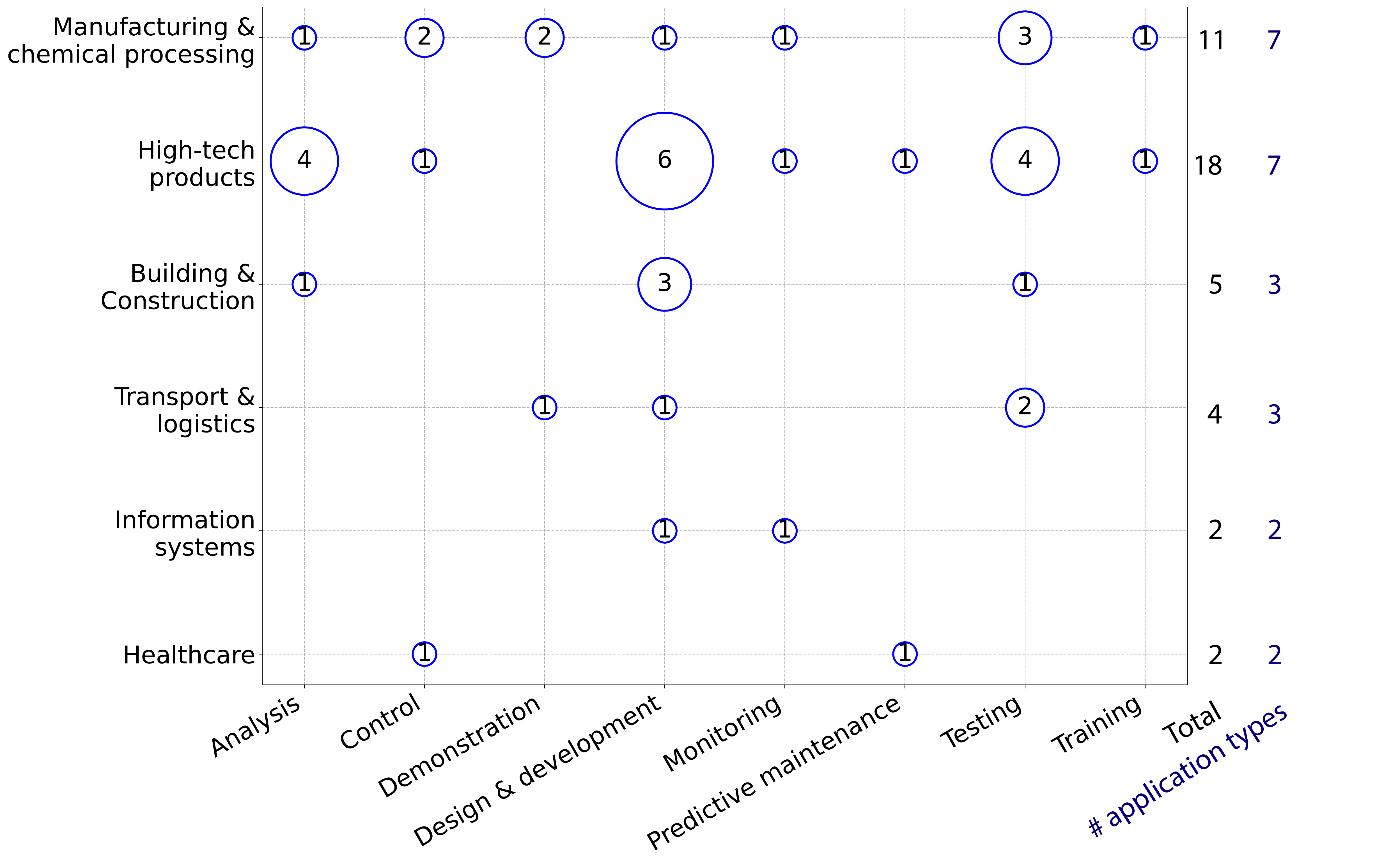}
    \caption{The correlation among the identified domains (y-axis) of the interviewees and DT's application (x-axis). The numbers on the plot represent the number of DT's applications in that specific domain. 
    }
    \label{fig:interviewees_domain_application}
\end{figure*}

\noindent Figure~\ref{fig:interviewees_domain_application} shows the correlation between the domains of the interviewees and the applications of their corresponding DTs. The \textit{total} value, shown in black (pre-final column), represents the sum of all the applications in a specific domain. The \textit{\# application types}, shown in blue (final column), represents the number of application categories of a specific domain.  The DTs in the high-tech products and manufacturing and chemical processing domains have the highest DT application diversity, each with seven application types. In these two domains, the most frequent DT applications are design and development, and testing. Furthermore, with the exception of the healthcare domain, all the other domains use DT for design and development. The other two most popular DT applications are testing---used by four domains---and analysis---used by three domains.

The findings presented in Figure~\ref{fig:interviewees_domain_application}  indicate that the prevailing DT applications among the interviewees are design and development, testing, and analysis. It is noteworthy that these DT applications align closely with the backgrounds of a substantial portion of the interviewees, which predominantly have a computer science brackground.

\section{Results and findings} \label{sec:results}
We now present the answers to our seven research questions. These answers are based on the analysis of the data, following the methodology described in \cref{sec:methodology}, collected during the interviews. Some of the discussions were not strictly related to the defined RQs but still yielded interesting insights; such additional findings are presented in \cref{sec:additional_findings_ext}.

\subsection{Definitions and understanding of DT (RQ1)}
\label{sec:def_understanding_DT}

As indicated it in Section~\ref{sec:introduction}, DT is used as an umbrella term: across different domains it
can have many different definitions, interpretations, and understandings. With RQ1 we aimed to understand these, and the similarities and differences between them.

A variety of definitions of DTs were discussed by the interviewees and there was no uniformity in these definitions.
Some interviewees have defined DTs with certain boundaries at the start of the interview, yet over the course of the interview mentioned additional aspects of DT, which extended their initial description. %
In Section~\ref{sec:virtual_entity},
we present views on DTs observed from the interview data.

The different components of a DT discussed during the interviews are presented in Table~\ref{tab:DT_components_table}, where the numbers indicate the number of interviewees from academia and industry respectively, who discussed those specific components of DTs. %
These numbers are out of the 10 participants from industry and 9 from academia as discussed in Section~\ref{sec:interviewee_demographics}; the reader should note that these numbers pertain as maxima for this table as well as for all later tables that refer to `\#I' (10) and `\#A' (9).

\subsubsection{Virtual representation of an entity} \label{sec:virtual_entity} 
We intended to find out how many interviewees agreed to the fundamental understanding of a DT being
a virtual representation of some entity. %
All the interviewees mentioned that a DT is
a virtual representation of some entity, by using different terms such as ‘digital counterpart’, ‘digital copy’, ‘virtual replica’, ‘virtual prototype’ or ‘model’. We believe that these terms mentioned by the interviewees refer to a model or a set of models, which we will define in Section~\ref{sec:DT_components}.  Five interviewees used terms such as ‘accurate’, ‘precise’ and ‘high fidelity’ to describe that a DT should be a high-fidelity representation of an entity. According to ten interviewees %
the level of fidelity is determined by the DT's purpose and application.
When further talking about DTs, interviewees discussed the type of entities the DT could virtually represent. These could be (1) a real world object with physical dimensions;
(2) a real world process or organization or even a concept without physical dimensions, such as a human resource process, a logistics process in manufacturing, 
fuzzy concepts 
and others.

Eight interviewees implicitly or explicitly discussed a DT being a virtual representation of not necessarily just a physical object, but of both physical and non-physical ones; indirectly referring to
a virtual representation of an \ActE{}.
Four interviewees did not explicitly mention whether DTs should be a virtual representation of an \ActE{}, but discussed their DTs being a virtual representation of a real world object with physical dimensions.
Ten interviewees also mentioned that DTs need not necessarily be a virtual representation of an existing \ActE{}, but could also be of an \ActE{} at the design stage. 
The current confusion in the description of DTs on whether the \ActE{} is part of the DT itself or not came up during the interviews.
Four interviewees expressed that it is not since the word ‘digital’ refers only to virtual objects and not physical objects. From the above, it can be understood that there is some level of alignment in the understanding of DT as a virtual representation of some entity which could be physical or non-physical, and which may or may not already exist.

\begin{table*}[ht]
    \centering
    \caption{Correlation between DT components from interviewees and 5D model~\cite{tao2017digital} addressed by industrial (\#I) and academic (\#A) interviewees.
    }
    \label{tab:DT_components_table}
    \begin{tabular}{c|c|cc}
        \hline
    \textbf{Interviewees' components} & \textbf{5D model components} & \textbf{\#I} & \textbf{\#A} \\ \hline
        Models & Virtual Entity & 10 & 9 \\
        Data & Data & 10 & 9 \\
        \ActE{} & Physical Entity & 4 & 8 \\
        Purpose & Services & 8 & 5 \\
        Uni-directional communication: \ActE{} $\rightarrow$ VE & Connections & 9 & 9 \\
        Bi-directional communication: \ActE{} $\leftrightarrow$ VE & Connections & 5 & 6 \\ \hline
    \end{tabular}
\end{table*}

\subsubsection{Components of a DT} \label{sec:DT_components} 
Through the interviews, we wanted to understand what components interviewees considered part of DTs; two questions were aimed towards this.
The different components of a DT discussed during the interviews are listed below. All these components of DTs are
represented in Table~\ref{tab:DT_components_table}, where the numbers represent the number of interviewees from academia and industry, respectively, who discussed those specific components of DTs.
\begin{itemize}

\item \textbf{Models:} Here, we consider models as 
an abstraction of the system which represents a certain viewpoint or aspects of the system such as its behavior, structure, function, performance or others. Interviewees 
mentioned different types of models, e.g.,~those representing geometry, physics, behavior and interactions; design and simulation models; descriptive and predictive models; 3D models; mathematical models; mechanical models; building information models (BIMs); CAD models; and others. All interviewees agreed explicitly or implicitly %
that models are an important component of DTs. 

\item \textbf{Data:} Interviewees %
also discussed different types of data such as data from sensor measurements; data from system; design data; historical data; reused data from the relevant product line which is in operation;
data acquired during DT operation; data from people who are part of the process; data from subject matter experts; data acquired during the entire lifecycle of a DT; and others. It can be concluded %
that all interviewees agreed explicitly or implicitly %
that data is an important component of DTs. 

\item \textbf{Purpose:} Thirteen interviewees expressed that a DT should have a purpose and some mentioned that this purpose is the driving factor for DTs to be developed. On the contrary, one industrial interviewee explicitly mentioned that DTs should {\em not} have a specific
purpose. This interviewee further discussed that DTs should not be developed with a specific
purpose and they have a purposeless existence, which is in stark contrast to what the thirteen interviewees mentioned above. This interviewee clarified that once the DT is developed, it can be used for whichever purpose is needed.
The purpose mentioned by the different interviewees can be correlated to the ‘services’ component of Tao et al.~\cite{tao2017digital}'s 5D model of DTs 
(see Section~\ref{sec:Background}).

\item \textbf{Communication between \ActE{} and its virtual counterpart:}

As part of the interview, we intended to understand the level of communication between the \ActE{} and its virtual counterpart. 
All but one interviewee discussed the synchronization from the former to the latter---either automated or manually.
 Such synchronization implicitly conveys unidirectional communication %
from \ActE{} to its corresponding virtual counterpart.
Eleven interviewees discussed that a DT should have bi-directional communication between the two entities. 
In addition, two interviewees also expressed that the virtual replica should not always be connected to its \ActE{} but only when this is needed. Five interviewees shared that the synchronization between \ActE{} and its virtual replica should be in real time. However, it  was not explicitly discussed further by any of the interviewees what was meant by `real time' which could possibly have different interpretations in different domains. 
We identified seven interviewees who mentioned that the frequency of synchronization depends on the purpose or application. %

\item \textbf{\ActE{}:} This component was already discussed in Section~\ref{sec:virtual_entity}.

\end{itemize}

\subsubsection{Relation to the 5D DT model}\label{sec:5D_perspective}

As explained in Section~\ref{sec:question_design}, we collected and analyzed the opinions of our interviewees on the 5D DT model by Tao et al.~\cite{tao2017digital} (explained in Section~\ref{sec:Background}). This 5D DT model was selected in this research because it is a widely used model for DT representation and has been adopted for discussions on DTs by many. So, we wanted to understand the opinions of interviewees on this highly used and adopted model. In this section, we present our observations.

We were able to map some of the DT components mentioned by the interviewees, as discussed in Section~\ref{sec:DT_components}, to the 5D model. 
This mapping is shown in Table~\ref{tab:DT_components_table}, where the number of interviewees from academia and industry who discussed the specific components of DTs is shown. %
As shown in that table, we identified considerable alignment
between the components mentioned by the interviewees and the physical entity, virtual entity, data and connection components of the 5D model. As discussed earlier, the purpose of a DT mentioned by different interviewees can be correlated to the services component in this model.

As depicted in Table~\ref{tab:Agreement_5Dmodel}, 11 interviewees mentioned that they could relate to the 5D DT model to a certain extent and agreed to this model albeit with some changes. Four interviewees from industry explicitly disagreed with this model. Another four interviewees neither explicitly agreed nor disagreed with this 5D DT model and discussed their perspectives on this model. It is also important to mention here that we were not able to obtain the opinion of one interviewee due to time constraints. %
Since we do not have data from this interviewee, we consider that this interviewee neither agrees nor disagrees to this 5D model.

\begin{table}[h!]
\centering
\caption{Agreement on 5D model of DT~\cite{tao2017digital} by interviewees from  academia (\#A) and industry (\#I).}
\label{tab:Agreement_5Dmodel}
\begin{tabular}{l|cc|c}
\hline
\multicolumn{1}{c|}{\textbf{Opinion on model}} & \textbf{\#I} & \textbf{\#A} & \textbf{Total} \\ \hline
Agree & 5 & 6 & 11 \\
Disagree & 4 & 0 & 4 \\
Neither & 1 & 3 & 4 \\ \hline
\end{tabular}
\end{table}

Three interviewees suggested that some connections in the 5D model are not needed and might be removed such as the connection between the PE and other components. They further elucidated that the connection from the PE might not be needed %
in some cases such as when the PE may not exist or when the PE may not be capable of communicating.
Two of those three further expressed that some of these connections need not be bi-directional, but can be uni-directional---such as the connection between data and PE, data and VE, and others. Some %
interviewees also mentioned that the nature of these connections were not clear enough. For example, one academic interviewee explicitly mentioned that this model should also clearly specify what data flows in each direction from one component to another. Some interviewees, albeit not significant numbers, also mentioned the role of humans in DTs and the prospect of DTs interacting through their services; these topics will be covered in \cref{sec:additional_findings_ext}. %

Overall, it can be concluded that although eleven interviewees agree to the 5D model, there are several changes suggested by them to this model and thus, they do not completely agree with it. While some of these changes do apply to DTs in general, some changes are also specific to the DT that the interviewee worked on.

\subsubsection{Discussion} \label{sec:Discussion_RQ1}
Based on our findings presented in this section, there is no common understanding of DTs across the nineteen interviewees. 
It is relevant to mention that this lack of common understanding has been discussed as one of the non-technical challenges in DT development by van den Brand et al.~\cite{Models_meet_data} and it is highly important to overcome this. 
Moreover, we aimed to understand the alignment of interviewees on the different parts that make up a DT, especially the highly relevant ones such as the relation and communication between \ActE{} and virtual counterpart.
It can also be observed from our findings that interviewees did not agree on all the components of a DT.

\takeaway{There is no uniformity in the definition of DTs nor in the understanding of the components that make up a DT. This lack of uniformity was also observed for the agreement on the 5D model of DT, even though it has been a widely used and adopted model of a DT. Despite this disparity, some agreement exists on certain components of a DT, specifically, models, data, and the synchronization between the AE and its virtual counterpart. 
Several changes were suggested to the 5D model of DT, such as removing some of the interconnections; having services interconnected with services from other DTs; adding humans as another dimension in DTs;
and others.
}

\subsection{Influence of reuse on DT lifecycle (RQ2)} \label{sec:reuse_rq2}

As indicated by Walravens et al.~\cite{walravensTurtle}, developing DTs is a cross-domain and resource intensive task where reuse of existing artifacts can significantly reduce the development and maintenance costs of DTs. With RQ2 our objective was to gain insight into the existing practices of reusing various artifacts for DT development. 
We identified 15 out of 19 interviewees acknowledging some form of artifact reuse while working with DTs. Industrial participants mentioned reuse more frequently.
Based on the interviews, we identified two kinds of artifact reuse: data reuse and software component reuse, which we discuss in \cref{sec:reuse_practices} and summarize in \cref{tab:reuse_type_summary}. We also discuss several related challenges that restrict the possibilities of reuse in \cref{sec:reuse_challenges}. Afterwards, in \cref{sec:discussion_rq2} we discuss our interpretation of these findings and what insights can be derived from them.

\begin{table*}
\centering
\caption{Mentions of artifact reuse by interviewees from industry (\#I) and academia (\#A).} 
\label{tab:reuse_type_summary}
\begin{tabular}{c|l|cc}
\hline
\textbf{Type of reuse} & \textbf{Types of reused artifacts}  & \textbf{\#I} & \textbf{\#A} \\ \hline
\multirow{5}{*}{Software}       & 3D/CAD models                 & \multirow{5}{*}{8} & \multirow{5}{*}{5} \\
                                & Design model & & \\ 
                                & Simulation model & & \\ 
                                & Software from product line & & \\ 
                                & Third-party (commercial) libraries & & \\ 
                                & Existing digital twin & & \\ \hline

\multirow{3}{*}{Data}           & Data (historical) from \ActE{}      & \multirow{3}{*}{5} & \multirow{3}{*}{2} \\ 
                                & Data from related product line & & \\
                                & Ontology & &\\ \hline

\multirow{2}{*}{Knowledge}      &    Experiences gathered at work   & \multirow{2}{*}{2} & \multirow{2}{*}{1} \\
                                &    Academic knowledge             & & \\ \hline
\end{tabular}
\end{table*}

\subsubsection{Current practices of reusing artifacts} \label{sec:reuse_practices}

\paragraph{Software artifact reuse} \label{sec:model_reuse}
We use this term to collectively identify the reuse of static and dynamic models and software components developed independently or extracted from one or more separate software intensive systems. Table~\ref{tab:reuse_type_summary} shows the types of reused software artifacts we identified through our analysis. 
We identified the following factors that encouraged or motivated software artifact reuse:
\begin{itemize}
    \item \textbf{Reduced resources for development:} Four interviewees mentioned that reusing software artifacts lead to shorter delivery time and reduced development effort. According to them, specially the ones from industry,
    reuse is essential as it greatly affects time to market. We additionally identified two cases where DTs were developed based on one or more existing DTs, allowing the developers to leverage existing artifacts and reduce development effort significantly. These identified benefits confirm our assumptions on the benefits of software artifact reuse based on prior publications reporting similar benefits for more traditional software systems~\cite{BARROSJUSTO20181, Mohagheghi2007}.
    
    \item \textbf{Ease of use:} According to six interviewees, ease of use and built-in support for integration provided by corresponding tooling encouraged them to reuse software artifacts. Based on our analysis, we divided these tools and integrated environments into two categories: \textit{commercial} and \textit{in-house tools}. While the commercially available tools are used both in industry and academia, the tools built in-house are exclusively used by the corresponding companies. Among the commercially available tools, \textit{Unity}~\cite{Unity_ref}, a well-known physics-based game engine, stood out as six interviewees mentioned to have used it or one of its commercial extension for developing geometry and physics models. Furthermore, our analysis suggests that in-house tools are often developed based on requirements defined by the organizations themselves and therefore are only suitable for their specific needs.

\end{itemize}

\paragraph{Data reuse} \label{sec:data_reuse}
Our analysis suggests that data reuse is an integral part of DT development. Seven interviewees explicitly mentioned it, three did so implicitly. Here we focus primarily on the explicit mentions. We found that the motivation for data reuse is related to the understandability of the data gathered from physical systems and its suitability for modeling certain aspects of the corresponding system into one or more DTs.
These DTs are later used for monitoring or enhancing the system. Although the nature, applications, and data sources for such reuse vary significantly, we identified two major types of data reuse:

\begin{itemize}
    \item \textbf{Data from similar systems:} Four interviewees mentioned developing DTs reusing data gathered from similar systems. %
    Three interviewees mentioned reusing data from a product line that is closely related to the DT under development. Our study suggests that this is particularly useful when the DT is being developed alongside a physical system that is not yet mature enough. 
    
    \item \textbf{Ontologies:} We found that both in industry and academia, the use of ontologies facilitates data reuse, especially when the data is produced and consumed in different contexts. In cases like this, ontologies are used to represent knowledge and describe various data properties. Three interviewees claimed that existing ontologies play an important role in their development of DTs. These interviewees are from the high-tech products and the building \& construction domain, which are known to be highly multidisciplinary.
\end{itemize}

\paragraph{Knowledge reuse}

We define knowledge as skills and experiences interviewees gather through education, training, and professional activities. It can be argued whether knowledge forms an artifact since it is not tangible and hard to measure. However, we identified three interviewees who emphasized the importance of preserving the experiences and practical knowledge gathered over time.

\subsubsection{Challenges in reusing artifacts} \label{sec:reuse_challenges}
Although fifteen interviewees mentioned practicing some form of artifact reuse and acknowledged its positive effects, we often identified cases where such reuse was restricted,
most frequently due to: 

\begin{itemize}
    \item \textbf{Legal issues:}
    We found that legal measures or clauses often restrict or prevent artifacts from being reused, especially when industrial stakeholders are involved. These measures include non-disclosure agreements (NDAs), intellectual property rights (IPR), and privacy concerns. Liability concerns can also have restrictive consequences, especially when multiple organizations are involved. With four interviewees explicitly mentioning it, we find this to be the most frequent challenge affecting both data and software reuse. 
    
    \item \textbf{Lack of explanation:} Our study suggests that the lack of appropriate description or documentation can severely reduce the possibility of artifact reuse. We found that data reuse is more affected by this. Three interviewees mentioned that various data is often collected to be used for specific purposes. Due to a lack of meta-data and knowledge lost over time, such data becomes meaningless, rendering reuse practically impossible. We identified that poorly documented or undocumented software components suffer from similar problems.

    \item \textbf{Incompatibility and integration issues:} This affects both software and data reuse. For the latter, this is often related to data formats being incompatible with available tooling. We also identified cases where precision and fidelity of available data restrict reuse. Software reuse also suffers from incompatibility issues that restrict the integration of existing components into newer systems. Lack of configurability, lack of interoperability between legacy and newer systems, and interface inconsistency are some of the factors that contribute to this issue. Our study also suggests that lack of proper documentation of software components can lead to perceived lack of interoperability.

    \item \textbf{Lack of methodology or tool:} This issue was identified by four interviewees as the reason behind limited reusability of existing artifacts. According to these interviewees identifying reusable components and determining the degree of reusability of the identified ones becomes very difficult due to inadequate consideration for future reusability during the development phase, and lack of appropriate methodology and mindset within the organization.
    We also found that most industrial DTs are developed using tools that are highly specialized and often built in-house. As a result, artifacts built using these tools are not easily reusable in a different context. 

    \item \textbf{Additional effort:} We identified four interviewees acknowledging the need for additional development and validation efforts to facilitate reuse of artifacts. The necessity to adapt existing software components for a new purpose is a major reason for this.
    Furthermore, newer operating conditions can reveal undetected defects of reused software components warranting further investigation and fixing.
\end{itemize}

\subsubsection{Discussion} \label{sec:discussion_rq2}

Based on the findings presented in this section, we observe that reusing existing artifacts can positively influence the development of DTs. We discussed reuse of software artifacts and data above. Besides these, the DT lifecycle can be significantly influenced by reuse of knowledge. However, we noticed that reuse is not practiced widely in the organizations that we interviewed from. In addition to the discussed reasons and challenges related to this reluctance, lesser mentioned ones include lack of appropriate software engineering skills, loss of knowledge due to personnel changes, and exceptional system requirements. These challenges and the ones listed above were mentioned by the interviewees in the context of DTs. However, we believe all of these challenges are relevant in the larger MDE context. Although resolving these might not be trivial (e.g., due to legal issues), we believe it is possible, with moderate organizational effort, to address issues related to inadequate tooling, lack of reuse attitude, and insufficient software engineering skills. Furthermore, during the interviews we noticed that organizations are recognizing the potential of reuse in the context of DTs and gradually moving towards developing, maintaining, and using reusable software components. We noticed a trend of developing modular or configurable DTs, often using a component-based integrated development environments for developing DTs by combining reusable components.

\takeaway{Reuse of existing data and software artifacts has the potential to significantly optimize the development lifecycle of a DT. However, except for some limited cases, it is not yet practiced widely due to challenges legal restrictions, inadequate tool support, lack of information, and lack of experience.
}

\subsection{Consistency management of cross-domain models (RQ3)} \label{sec:consistency}

Cross-domain collaboration is essential for the development and maintenance of DTs. Within a cross-domain environment, we identified maintaining consistency among cross-domain software models as a challenge~\cite{Models_meet_data}. With RQ3 we aimed to investigate this challenge.
We wanted to understand how consistency is defined in practice; and identify key causes for inconsistencies, and tools and techniques used to manage them. Twelve interviewees mentioned that they have encountered or put measures in place to handle inconsistencies. In the following sections we present our findings based on the analysis of the data we collected from these interviewees.

\begin{table}[ht]
    \centering
    \caption{Various inconsistencies encountered by interviewees from industry (\#I) and academia (\#A).}
    \label{tab:encountered_inconsistency}
    \begin{tabular}{l|cc}
        \hline
        \textbf{Encountered inconsistencies} & \textbf{\#I} & \textbf{\#A}\\ \hline
        Interface inconsistency & 2 & 2 \\
        Model format inconsistency & 3 & 3 \\
        Representation inconsistency & 1 & 2 \\
        \ActE{}-VE inconsistency & 2 & 1 \\ \hline
    \end{tabular}
\end{table}

\subsubsection{Inconsistencies encountered in practice}\label{sub:inconsistency-type}

Based on our analysis, we identified several types of inconsistencies encountered in practice. We discuss the four types that appeared most often during the interviews. \cref{tab:encountered_inconsistency} shows an overview with each of the inconsistency types and the number of interviewees mentioning them.

\begin{itemize}
    \item \textbf{Interface inconsistency:} Hisarciklilar et al.~\cite{Hisarciklilar2011} defined interface inconsistency as mismatching values, terminologies, or schemes among connected interface elements. We identified this inconsistency mostly in cases where two or more models, often cross-domain, need to communicate and do not share any compatible interfaces. Our analysis suggests that this is one of the most frequently encountered inconsistencies, with four interviewees explicitly mentioning it.

    \item \textbf{Model format inconsistency:} Six interviewees recognized such inconsistencies as a challenge. Our analysis shows that the development of DTs is almost always a cross-domain effort. In projects such as these, the stakeholders are from a variety of domains and use domain-specific tooling to develop cross-domain models and artifacts. From the interviews, we identified situations where these tools are completely or partially incompatible. As a result, models developed in one tool can not be imported to or used in another tool, primarily due to incompatible formats.

    \item \textbf{Design-implementation inconsistency:} As DTs often represent complex cyber-physical systems, various diagrams (i.e., UML~\cite{om2017unified}, SysML~\cite{hause2006sysml}) are used to conceptually represent parts or the complete system, usually during the design phase. In our analysis we identified cases where the actual implementation deviates from the design, which we identify as representation inconsistency.
    
    \item \textbf{\ActE{}-VE inconsistency:}
    As discussed in Section~\ref{sec:def_understanding_DT}, the concept of a DT often encompasses a certain degree of synchronization between \ActE{} and the corresponding VE to facilitate mimicking certain behavior or features. We identified two types of inconsistency in this context. The first kind is about inconsistent \ActE{}-VE communication, which is often a special kind of interface inconsistency (discussed earlier). The other inconsistency occurs when the collective behavior of models within the VE and the behavior of the \ActE{} are too dissimilar to be fit for the services the DT needs to offer. As a result, identical operations performed on both \ActE{} and VE can provide different results, rendering the DT ineffective. 
\end{itemize}

\subsubsection{Challenges in maintaining consistency}
\label{sec:reason_inconsistency}

Our analysis of the interviews demonstrates that maintaining consistency among various artifacts in the context of DTs is not a trivial task; interviewees mentioned various challenges they encounter while trying to do so. In this section we discuss the most prominent challenges we identified. Although this discussion includes a limited number of challenges and existence of more is highly likely, our finding already shows the complexity of maintaining a consistent system.

With five interviewees explicitly mentioning lack of standardisation, we identified this as one of the most frequent reasons for both model and data related inconsistencies. Two of them mentioned that they are not aware of any standardisation within their project, resulting in significant overhead in terms of communication and development efforts. We also identified cases where data was collected, stored, and exchanged using a non-standard format despite the existence of established standards. It was unclear from the interviews why existing standards were not followed. Furthermore, one of the interviewees mentioned that developing and following a set of standards for exchanging or storing information is extremely challenging due to the sheer number of different domains involved in their DT project.

We also identified lack of proper collaboration, insufficient tooling or methodology, and reuse of artifacts as challenges. Although each of these were mentioned only by one or two interviewees, they are potentially broad in impact.

\subsubsection{Inconsistency mitigation practices} \label{sec:inconsistency_mitigation}
During the interviews, we tried to understand how inconsistencies are being handled in practice in the context of DTs. Below we discuss the most prominent of the measures we identified: 

\begin{itemize}
    \item \textbf{Use of standards:} We found that the usage of standards is one of the key practices for avoiding inconsistencies and maintaining consistency. Our analysis suggests that these standards can be globally accepted or custom-built for an organization. Five interviewees mentioned using one or more of the following standards: Functional Mock-up Interface (FMI), Functional Mock-up Unit (FMU), European Union (EU) defined standard (no further details were provided), and organization-specific customized standard.

    \item \textbf{Use of external tool or technology:} Four interviewees mentioned using external tools for handling inconsistency issues. In this context, we found that semantic technologies play a key role in understanding of data and, in some cases, conversion between different data formats. These interviewees mentioned the usage of ontologies and related technologies, i.e., SHACL\footnote{Shapes Constraint Language \url{https://www.w3.org/TR/shacl}}, and graph databases, i.e., Neo4j\footnote{Neo4J - a graph data platform \url{https://neo4j.com}}.

    \item \textbf{Formal or informal communication:} Earlier we mentioned general lack of tooling or methodology as one of the reasons for inconsistency. In fact, we believe a majority of inconsistencies are avoided by established practices within an organization, way of working, in-person communication, or personal knowledge. Three interviewees indicated that they often needed to resort to informal in-person communication to resolve inconsistency issues. Our analysis shows that such communication can take place within or across organizations and domain.
    
    \item \textbf{Use of in-house tooling:} We discussed software related inconsistencies in the context of multi-tool environments in Section~\ref{sub:inconsistency-type}. Our analysis suggests that one of the practices for reducing the number of inter-tool inconsistencies is to avoid multiple tools and using a single one. Our analysis identified three individual cases where this approach was employed.

    \item \textbf{Testing:} In safety critical domains, e.g., aerospace and healthcare, early and frequent testing or benchmarking was mentioned as a strategy to identify potential problems including inconsistencies.

\end{itemize}

\subsubsection{Discussion}

Our analysis suggests that inconsistencies are actual issues in the context of DTs and directly or indirectly affected over 60\% of our interviewees. The nature and source of these issues are highly diverse and depend on the actual DT implementation, involved methods and tooling, organizational and personal constraints. Consequently, we believe that it is fairly impossible to categorise these inconsistencies completely. Furthermore, we identified that inconsistency issues are often not categorized as such. Instead, they are treated as regular issues encountered during system development or maintenance. Therefore, we believe there is a lack of awareness of the need for specialized methodologies or techniques for identifying and mitigating inconsistencies. We think this might have contributed to the inadequacy of related tooling, as discussed before.
Furthermore, we identified situations where the vast majority of inconsistency issues are being avoided by simply using a single tool for development, or by developing large monolithic models. Besides, we think there is a large gap between academic innovation and industrial practice in the context of inconsistency management. For example, Torres et al.~\cite{torres2020systematic} referred to several academically developed consistency management approaches in their systematic literature review. However, we could not identify any commonality between this list and the approaches discussed by our interviewees, which we present in Section~\ref{sec:inconsistency_mitigation}. Therefore, we believe that there are opportunities for further research and development for consistency management tools in the context of DTs. However, we also identified that certain domains, i.e., aerospace, construction engineering, automotive, are more aware of and mature about handling inconsistency issues. This is largely facilitated by safety requirements and standards or conventions accepted across related organizational entities.

\takeaway{In the context of DTs, inconsistency issues are common and can adversely affect development and maintenance activities. We identified appropriate communication and usage of standards as the most frequently practiced measures to avoid inconsistencies. Contrarily, we noticed that the absence of these measures are the major reasons behind the emergence of inconsistency issues. Furthermore, we observed a general lack of tooling and methodology for effectively handling consistency. Therefore, we believe that further research, development, and adoption is needed to understand these issues and develop tools and techniques to avoid or mitigate them.}

\subsection{Model integration (RQ4)}\label{sec:integration_Rq4}

RQ4 explores the topic of model integration in DTs. Model integration is the process of bringing together models to create the DT virtual entity. These models will interact to mimic a desired behavior from the \ActE{}. We intended to understand the approaches and design decisions the interviewees take when designing a DT. 

\subsubsection{Model integration practices}
Regarding model integration, an overview of the interviewees' approaches and design decisions are shown in \cref{tab:rq4Overview-integration}. The second column shows our classification of the findings. 

\begin{table*}[h!]
\centering
\caption{Overview of model integration discussed by industrial (\#I) and academic (\#A) interviewees.}
\label{tab:rq4Overview-integration}
\begin{tabular}{c|l|cc}
\hline
\textbf{Topics discussed}                            & \multicolumn{1}{c|}{\textbf{Findings classification}} & \textbf{\#I} & \textbf{\#A} \\ \hline
\multirow{2}{*}{Approach}                   & Multi-tool                          & 6   & 3   \\
                                            & Single-tool                         & 4   & 1   \\ \hline
\multirow{2}{*}{Communication among models} & Design considerations               & 5   & 4   \\
                                            & Implementation approach             & 5   & 1   \\ \hline
\multirow{4}{*}{Technology used}            & Functional Mock-up Interface (FMI)  & 3   & 0   \\
                                            & Domain Specific Language (DSL)      & 1   & 2   \\
                                            & Own-design                          & 4   & 0   \\
                                            & Tool provided                       & 2   & 2   \\ \hline
\multirow{2}{*}{Tooling type}               & In-house                            & 7   & 1   \\
                                            & Commercial                          & 2   & 4   \\ \hline
\end{tabular}
\end{table*}

\paragraph{Integration approaches}
\label{subsubsection:integration_type}

From our analysis, we observed two main integration approaches, namely multi-tool (nine interviewees) and single-tool (five interviewees). %
\begin{itemize}
    \item \textbf{Multi-tool approach}. This approach is used when a heterogeneous modeling environment is present, in which distinct modeling tools are combined~\cite{Ptolemaeus:14:SystemDesign}. Each model needs encapsulation resulting in a defined interface to establish communication with other models. The nine interviewees using the multi-tool approach agree that this approach has advantages for cross-domain collaboration because it allows different tools to be used. Other reasons mentioned to use this approach is information hiding of model details also known as model masking, for IP %
    protection and reduction of model complexity. Here, technologies are used so the model becomes a ``black box'' with only its input and output exposed, and hiding the model details. However, this approach has many challenges such as interface consistency (discussed in~\cref{sub:inconsistency-type}), data and model semantics, and relationship complexity between the models. 
    \item \textbf{Single-tool approach}. This approach requires to generate or transform all the models for use with a single software tool. The single tool will perform the execution of all the models, i.e. serves as execution platform. Our analysis showed that the selected execution platform in such cases was MATLAB. In addition, our analysis shows two strategies from interviewees. The first strategy is to use the same tool which is used for model execution for creating the models as well, e.g., two interviewees use MATLAB as their modeling environment and execution platform. The second strategy is to transform the original models into a format which is supported by the execution platform. Interviewees using this strategy mentioned that this requires re-work from them. An example of this practice is the use of Python as an execution platform that can support the execution of models made in Python or MATLAB. If there is a model in Modelica then this model is transformed into MATLAB, the supported platform, requiring extra work from the modeler. Interviewees mentioned that this approach might limit cross-domain collaboration, but significantly reduces the integration effort.  
\end{itemize}

\paragraph{Communication among models}
As mentioned previously, an important ingredient of model integration is the communication between models. The communication between models refers to the action of transmitting data between models, particularly between models built in different modeling tools. We aimed to understand how the interviewees implemented such communication.  We identified 15 interviewees who addressed this topic. Our analysis showed two key topics mentioned: %

\begin{enumerate}
    \item \textbf{Important factors that influence communication design}. We identified three factors influencing model communication, namely the DT's application, software dependencies, and stakeholders' involvement. According to nine interviewees, the DT's application dictates the required communication frequency or the implementation approach. However, they mentioned that stakeholders' involvement is crucial because it will determine the implementation feasibility by providing supporting knowledge, resources, and software. Moreover, they mentioned the importance of knowing the dependencies and requirements of the software modeling tools or platforms involved, to avoid operational failure.
    \item \textbf{Implementation approach}. We identified two main approaches to implement the communication, namely use of standardized communication protocols and in-house technology. Five interviewees shared that they use standardized communication protocols. Three of them use OPC~\footnote{\url{https://opcfoundation.org/}} (Open Platform Communication). The second approach used by remaining is an in-house technology similar to a publish-and-subscribe pattern for data exchange.

\end{enumerate}

\noindent In conclusion, our analysis shows that the most popular implementation approach is the use of communication protocols, used by five out of the six interviewees who addressed this topic.

\paragraph{Technology used}

During the interviews, we aimed %
to understand the type of technologies used for model integration. Fourteen interviewees described different technologies that are used for this purpose.  
Table~\ref{tab:rq4Overview-integration} shows a summary of this subsection, where the technologies used varied greatly, but can be clustered in four groups: 
Functional Mock-up Interface (FMI), DSL, own-design and tool-provided technologies.

\begin{itemize}
    \item \textbf{FMI}: Three interviewees specifically mentioned the use of FMI technology to integrate their models. This technology has been used in DTs~\cite{FMU_in_DTs} and is supported by over 170 modeling tools\footnote{\url{https://fmi-standard.org/}}. 
    According to three interviewees from industry, 
    two reasons to use this technology are its maturity and compatibility with various modeling tools. 
    \item \textbf{DSL}: Three interviewees mentioned the use of DSLs for model encapsulation and for establishing communication between these models. Interviewees also use this technology for other aspects such as DT architecture (see Section~\ref{sec:additional_findings_ext}) and to unify data semantics among the models. 
    \item \textbf{Own-design}: Four industrial interviewees indicated that they developed their own technology to integrate models in their DTs. Their technology is based on developing the interfaces for each modeling tool they have used, e.g., if the interviewees have models in the MATLAB and ANSYS modeling tools, they design an interface for each of them. According to these interviewees, this method gives them flexibility, and it can be expanded according to their needs. However, they admit that it requires effort and time every time a new modeling tool is added.
        
    \item \textbf{Tool provided}: Another four interviewees mentioned that they use what is supported by their execution platform. As with the monolithic approach explained in Section~\ref{subsubsection:integration_type}, these interviewees have two choices: either transform incompatible models or develop them in a format supported by their execution platform. According to these interviewees, the main reason to choose the technology is because of their experience with the execution platform.
  
\end{itemize}

\paragraph{Tooling type} \label{sub:integration_tool}
The interviewees mentioned two distinct uses of tooling. First, tools to develop models for their DTs, for which they all mentioned using commercial software such as MATLAB. Second, tools to execute all the models, as discussed in  Section~\ref{sec:integration_Rq4}. We divided the identified execution platforms into two categories: developed in-house and commercial. Table~\ref{tab:rq4Overview-integration} shows a preference for the development of in-house tooling among the interviewees, particularly industrial interviewees. %
Six interviewees did not mention the tooling used for DT integration.  

\begin{itemize}
    \item \textbf{In-house}: The data collected suggests that interviewees from industry prefer to develop their own tools for DT development. The main driver to do so seems to be the DT's application and its domain requirements. Two interviewees mentioned the need for specific execution and modeling requirements for their event-based DTs, driven by the DT application needs. Another two based their tooling on visual models, which required specific communication technologies and the integration with specific software tooling from their stakeholders. The final four interviewees mentioned that they use modeling tools (commercial and in-house) frequently used in their own domain and thus required to work together. This resulted in the need to create their own tool for integrating models. To summarize, industrial interviewees seem to prefer to develop their own tools to realize their specific integration requirements. 

    \item  \textbf{Commercial}: The use of commercial tools varies between modeling (e.g., MATLAB), CI/CD (Continuous Integration/Continuous Development), 
    cloud frameworks, and system design tools. According to the interviewees, the main reason for their use is the efficiency of component integration. However, commercial tools have limited support for external software. 
    As a consequence, these interviewees are forced to transform their models to a supported format; this might imply re-work or limit collaboration.
\end{itemize}

\subsubsection{Challenges in model integration}\label{subsec:integration_challenges}
Seven interviewees shared the challenges they face when integrating models for DTs. Based on our analysis we decided to classify these challenges in three groups, namely challenges in model heterogeneity, data heterogeneity, and complexity.

\begin{itemize}
    \item \textbf{Model Heterogeneity:} These challenges are related to the types of models that need to be integrated into the DT. We identified two challenges related to models: integration of legacy models, extracted from legacy code, and integration of models from different software platforms. Three interviewees mentioned that these challenges seem to be particularly difficult and require further research to address them. 
    \item \textbf{Data Heterogeneity:} The data heterogeneity challenges refer to data format and semantics; both challenges are present due to cross-domain collaboration. An example of a semantics challenge is when two terms referring to the same concept are used in different domains, such as ``pressure drop'' and ``pressure gradient''. Different terminology can generate confusion which might impact the development. According to our analysis, interviewees seem to find ways to address this challenge. For the data format challenge, a solution is the development of a communication layer to homologize data formats between models. For the semantics challenge, an interviewee uses semantic web technology to standardize the semantics. 
    \item \textbf{Complexity:}The complexity challenges are related to models and the DT as a whole. Two interviewees defined complexity of a model as the level of fidelity. These interviewees suggest considering the purpose of the DT as the key factor for design. Therefore, the critical aspect lies in choosing the appropriate level of model fidelity to prevent the need for integration rework. The second complexity challenge discussed is related to the system as a whole. Two interviewees stated that a DT can be composed of several components, increasing the complexity, and hence leading to difficulties in understanding the relation between the components. In addition, understanding those relations becomes critical to solve integration issues. 

\end{itemize}

\subsubsection{Discussion} 

Based on our analysis the multi-tool approach for integration seems to be most popular among the interviewees. Our opinion is that the multi-tool approach offers better maintainability and cross-domain cooperation, due to the separation of entities. On the other hand, interviewees agreed that this approach requires more effort and knowledge of software engineering. Among the interviewees, the most popular approach for model communication is using communication protocols such as OPC. According to our analysis, the selection of communication seems to be influenced largely by the experience of the developer. Through the interviewees, we found diverse technologies used for integration. However, two technologies were mentioned by three interviewees each: DSLs and FMI. Regarding tooling for integration, seven out of nine industrial interviewees prefer in-house tooling. Our analysis of the challenges suggests that tooling and technologies to facilitate cross-platform integration are required.

\takeaway{%
        The preferred approach for the integration of models is a multi-tool approach which requires interface development. The preferred technology for such interfaces seems to be the use of standardized communication protocols. Although there is no clear preference for integration technology, two technologies seem to be frequently used, FMI and DSL. Finally, integration has three main challenges: model heterogeneity, data heterogeneity, and DT's complexity.
        }

\subsection{Model orchestration (RQ5)}\label{sec:orchestration_rq5}

We define DT model orchestration as the definition of the required communication actions and correct execution sequence of the models~\cite{Standard_orchestration}. To achieve this the orchestration should include activities such as data compatibility checks and starting and ending model execution~\cite{BP_heterogeneousSys}. With RQ5 we aimed to understand interviewees' perceptions and practices related to such orchestration in DTs. 

\subsubsection{Model orchestration practices}

We identified five main topics of discussion related to model orchestration, namely understanding, implementation, technology, tools, and challenges. An overview of those topics is shown in Table~\ref{tab:orchestration_results}'s first column. The second column shows our classification of the findings to facilitate reading the results. The third column shows the number of interviewees who discussed an item from a specific class. 

\begin{table*}[h!]
\centering
\caption{Overview of model orchestration discussed by industrial (\#I) and academic (\#A) interviewees. }
\label{tab:orchestration_results}
\begin{tabular}{c|l|cc}
\hline
\textbf{Topics   discussed}                       & \multicolumn{1}{c|}{\textbf{Findings classification}} & \textbf{\#I} & \textbf{\#A} \\ \hline
\multirow{2}{*}{Description}             & Definition                                   & 9   & 2   \\
                                         & Components                                   & 4   & 1   \\ \hline
\multirow{2}{*}{Implementation approach} & Pragmatic                                    & 2   & 4   \\
                                         & DT's Application specific                    & 6   & 1   \\ \hline
\multirow{2}{*}{Technology}              & Model based                                  & 3   & 2   \\
                                         & Own design                                   & 3   & 0   \\
                                          \hline
\multirow{2}{*}{Tools}                   & In-house                                     & 6   & 0   \\
                                         & Commercial                                   & 2   & 5   \\ \hline
\end{tabular}
\end{table*}

\paragraph{Understanding}\label{subsub:orch_description}

We divided the discussion into two topics: interviewees' explanation of what orchestration is, and of its components.

\begin{itemize}
    \item \textbf{Definition}. Eleven interviewees shared their definitions of model orchestration. They all agree that orchestration is the scheduling of model execution in their DTs. Only four of them specifically mentioned that the method of data exchange is part of the orchestration. In addition, these eleven interviewees also expressed their opinion on the importance of orchestration. From that we concluded that model orchestration is highly important, as three interviewees explicitly expressed it and another six implicitly did so. Yet two interviewees argued that orchestration is not needed in their DTs, because the complexity of their current DTs is not high. %
    \item \textbf{Components}.  Five interviewees specifically shared the necessary components to design the orchestration of models. All other interviewees mentioned that their orchestration implementation is dependent on purpose and domain; thus they did not define specific components for orchestration. Table~\ref{tab:orchestration_components} shows the components mentioned by the five aforementioned interviewees and the number of mentions for each component. Concerning the trigger, eight interviewees explicitly stated that it is a key component of orchestration. Nevertheless, the type of trigger depends on the DT's application and domain. We observed two distinctive trigger definitions as a function of the DT's application. Two interviewees working on control applications stated that the orchestration should be done based on a time schedule, where the data exchange between models and the execution of each model should be synchronized based on a global clock. Another interviewee, working on event-based applications, mentioned that the definition of the trigger for each event is the most important factor to schedule each model execution step. The remaining five interviewees explained that the trigger for model execution depends on DT's application and domain.

Regarding the scheduling approach, interviewees mentioned two types, namely sequential scheduling and concurrent execution. Regarding the data exchange method, interviewees defined it as the scheduling of data exchange between models, e.g., First In, First Out. Our analysis shows two different roles of time in orchestration. The first role is as a trigger for model execution in control applications, known as time-based scheduling. The second role is event recording in event-based applications, by using time stamps for each event (cf.~Table~\ref{tab:orchestration_components}). 
\end{itemize}

\begin{table}[h!]
\caption{Components to define orchestration in DTs.}
\label{tab:orchestration_components}
\centering
\begin{tabular}{l|c}
\hline
\multicolumn{1}{c}{\textbf{Components}}              & \multicolumn{1}{l}{\textbf{Mentions}} \\ \hline
Trigger                 & 8                            \\
Scheduling approach     & 4                            \\
Data exchange method    & 4                            \\
Global time             & 3                            \\
Time-stamping             & 3                            \\\hline
\end{tabular}
\end{table}

\noindent In conclusion, our analysis shows a general consensus on orchestration as all activities ensuring correct scheduling of model execution. The majority of the interviewees consider orchestration important for the development of DTs. These interviewees also agree that the orchestration design requires a definition of the scheduling and method for data exchange. In addition, they agree to define a global time for the DT application and labeling produced data with time-stamps. Other components for the orchestration design seem to depend on the DT's applications and domain.

\paragraph{Implementation approach}\label{subsub:orch_implementation}
Thirteen interviewees shared their specific implementation method. Six interviewees stated to use a pragmatic approach, while the other seven shared specific implementation approaches, depending on their domain.
       
\begin{figure*}[h!]
    \centering
    \includegraphics[width=0.75\textwidth]{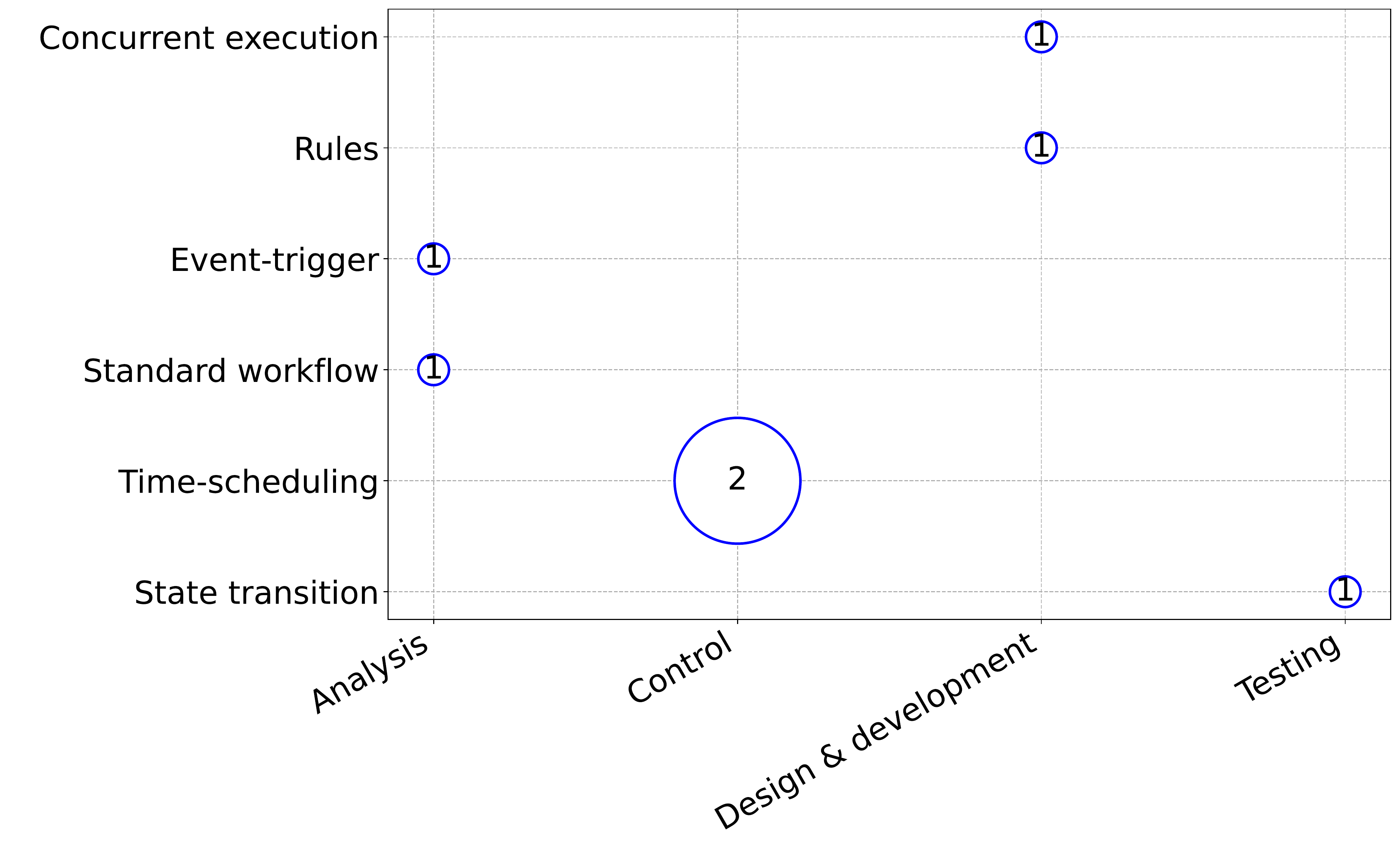}
    \caption{DT's application specific approaches for model orchestration. 
    }
    \label{fig:approach_orchestration}
\end{figure*}

\begin{itemize}
    \item \textbf{Pragmatic}.  Our analysis suggests that the pragmatic approach aims to design the scheduling of models by reproducing the \ActE{} behavior, e.g., if two tasks occur simultaneously, then concurrent execution is used. According to the interviewees, this approach requires iterative testing, design, and implementation. Our analysis shows that interviewees using the pragmatic approach design the orchestration by directly writing code to define the scheduling of models. 

    \item \textbf{DT application specific}.  We found seven interviewees who stated that the DT application determines the orchestration approach. During our interviews, we collected six different approaches, shown in Figure~\ref{fig:approach_orchestration}. Each approach defines how to implement the models' scheduling. The only approach used by multiple interviewees, with control applications, is time-scheduling, in which time triggers execution for each model. In addition to the DT application, the orchestration approach seems to be related to the knowledge and domain of the interviewee. \cref{fig:approach_orchestration} shows two examples. The first example is related to Design \& development, which has two distinct approaches 1) rules, and 2) concurrent execution; these approaches are chosen based on their knowledge. The second example is on Analysis, which also has two approaches: 1) standard workflow, and 2) event-trigger; chosen based on the domain.
\end{itemize}

\noindent In conclusion, the interviews indicate that the approaches are a function of the DT's application, and of interviewee knowledge and domain. In addition, around half of the interviewees seem to design the orchestration by attempting to pragmatically reproduce the \ActE{} behavior.

\paragraph{Technology used}\label{subsub:orch_tech}
Ten interviewees shared the specific technologies they used for orchestration. We have classified these into two categories: model-based and own design.

\begin{itemize}
    \item \textbf{Model-based}. Five interviewees stated their preference in using technologies that are model-based to design the orchestration. The two technologies described by these interviewees were DSLs and ontologies. Four interviewees use DSLs to design the scheduling of the models. Three of them defined their own DSL, while another uses SysML. One of them also uses their DSL for system verification. Another interviewee uses an ontology to link data between models and thereby orchestrate data exchange.

    \item \textbf{Own design}. Three industrial interviewees explained that they designed their own technology for orchestration. The technology is based on their expertise and domain. None of them explained their technology in detail but rather shared how it works at a higher abstraction level.   We identified two distinctive technologies based on the trigger type, i.e., time or event-based triggers.

\end{itemize}

In conclusion, our analysis suggests that the most popular technology is model-based, particularly DSLs. We observe that all orchestration technologies focus on the scheduling of the models, but not on how to exchange data among them.  

\paragraph{Tool type}\label{subsub:orch_tool}

From 19 interviewees 13 shared the tools they use, in particular their execution platform, which performs the orchestration in a DT. Six interviewees from industry developed their own tool, while seven (five from academia and two from industry) use a commercial tool. %

\begin{itemize}
    \item \textbf{In-house}. Six interviewees explained that they have developed their own execution platform to schedule model execution. Based on our findings, we observe two main reasons for in-house tool development: (1) to minimize integration effort, for example use C\# to create models and as an execution platform; and (2) to support specific execution requirements, such as specific execution time requirements, like simulate a model with specific time configurations.

    \item \textbf{Commercial}. We identified three types of commercial tools used as execution platforms. The first is to use modeling tools to orchestrate the models, e.g., MATLAB. This type of tool is used by two interviewees who also use the single-tool approach for the integration of models as explained in Section~\ref{subsubsection:integration_type}. The second type is the use of system design tools such as IBM Rhapsody~\cite{IBMUserGuide} and HEEDS\footnote{\url{https://www.plm.automation.siemens.com/global/en/products/simcenter/simcenter-heeds.html}}, used by two interviewees. These tools facilitate the use of external software as long as they are supported by the vendor. The third type encompasses tools that support a DSL to sequence entities for execution; this tool type is used by two interviewees. The tools mentioned by interviewees are PDDL (Planning Domain Definition Language) and Dezyne from Verum\footnote{\url{https://www.verum.com/DiscoverDezyne}}. The latter also supports formal verification.
       
\end{itemize}

\noindent Based on our findings, we observed that the use of commercial tools is slightly preferred over the development of in-house tools. Interviewees did not specifically mention why they preferred commercial tools, but three implied some reasons: previous tool knowledge, external software support, and facilitating DT system integration. The last reason was indicated by the interviewees who use a modeling tool for orchestration because they considered it easier to transform all models into a single modeling tool than to orchestrate cross-platform models. 

\subsubsection{Challenges in model orchestration}\label{subsub:orch_challenge}
Through the interviews, different challenges were mentioned by interviewees. We have classified them in two groups: model fidelity and interoperability. From 19 interviewees nine mentioned such challenges. 
\begin{itemize}
    \item \textbf{Model fidelity:} %
    We found three interviewees addressing this challenge. The fidelity of a model is defined by interviewees as the level of accuracy between the \ActE{} and its model. This challenge particularly deals with conflicting requirements between real-time execution and high-fidelity models. %

    \item \textbf{Interoperability:} The main challenge discussed by four interviewees, is related to the cross-platform and heterogeneous nature of models for DTs, which yields semantic challenges. 
\end{itemize}

\subsubsection{Discussion} 
Our analysis suggests that the majority of interviewees agree that orchestration is to correctly schedule models' execution. In addition, there are five key components to implement the orchestration: a trigger for model execution, a scheduling approach, data exchange method, global time, and time-stamps as shown in Table~\ref{tab:orchestration_components}. We observed various scheduling approaches that are highly influenced by the DT's application and developers' knowledge. We believe that to facilitate orchestration design, more research should be done to create a general approach.

Technology selection for orchestration seems to favor model-based approaches, with DSLs as the most popular. Our findings suggest that influencing factors for tool selection are the interviewee's domain and previous knowledge of specific tools.

We believe that the challenges related to model fidelity and model understanding can be tackled by clearly defining a DT's purpose and developing or modifying the models accordingly. Regarding the interoperability challenge, we believe that research on tools to facilitate interoperability, particularly in cross-platform and model-type interoperability can tackle this challenge.

\takeaway{The main task of orchestration is to correctly schedule models' execution and a key component is the execution trigger. %
The orchestration design seems to require much domain knowledge and is highly influenced by the DT's application and the developers' knowledge. Technology and tool selection is also highly influenced by DT's application. Further research is needed on tools to facilitate the interoperability of cross-platform and cross-nature model types.}

\subsection{Validation and verification techniques and tools (RQ6)}
\label{sec:validation-techniques-tools}

RQ6 aims to understand what specific techniques and tools are used to verify and validate DTs and their overall dynamic behavior, that is, 
the behavior observed during the collective execution of the models and other components in VE. We identified 18 out of the 19 interviewees validating their DT.
It was observed that when answering the related interview questions, interviewees tended to use ‘system’ to interchangeably describe either the DT or the VE or the \ActE{}.
We found three different techniques used by the 18 interviewees for validating their DT, namely by comparing the AE and the VE behaviour, use of formal methods, and testing. 

These techniques are discussed in detail in Section~\ref{sec:validation_techniques}. Moreover, interviewees discussed the different challenges involved with verification and validation of DTs which we cover in Section~\ref{sec:challenges_validation}. 
Table~\ref{tab:VV_techniques} summarizes the different verification and validation techniques and strategies put forward by interviewees.
\begin{table*}[ht]
\centering
\caption{Different verification and validation techniques and strategies used by interviewees from industry (\#I) and academia (\#A).}
\label{tab:VV_techniques}
\begin{tabular}{l|l|cc}
\hline
\multicolumn{1}{c|}{\textbf{Topic}} & \multicolumn{1}{c|}{\textbf{Findings}} & \textbf{\#I} & \textbf{\#A} \\ \hline
\multirow{2}{*}{Techniques} & Comparing AE and VE behaviour & 8 & 5 \\
 & Formal Methods & 2 & 1 \\ \hline
\multirow{2}{*}{Strategies} & Increase complexity gradually & 2 & 1 \\
 & Continuous validation & 2 & 1 \\ \hline
\end{tabular}
\end{table*}

\subsubsection{Verification and validation techniques}\label{sec:validation_techniques}%
We believe validation of DTs to be highly important, as it reduces the possibility of errors while executing. 
Furthermore, our analysis suggests that validation is required in cases where a highly accurate
DT is being used. Such highly accurate DTs of critical modules are then reused for different purposes across product lines, hence, across multiple DTs.

The observations from the interview may not necessarily encompass all aspects of DTs which need to be validated, but only those the interviewees stumbled upon in their DT or which they consider of highest importance from their standpoint.
We observed that validation of DTs is highly dependent on their purpose, the DT's application domain, and the types of models and data used in the DT. For example, an interviewee from industry mentioned that when DTs are created as visualizations for marketing, validation is not required. However, he also mentioned that
when high fidelity, consistent behavior, and reliability in DTs are required, validation is crucial. Interviewees discussed various techniques for verification and validation based on different cases as depicted in Table~\ref{tab:VV_techniques}; we next present our findings on these. %

\begin{itemize}
{
\item \textbf{Comparing \ActE{} and VE behavior}
\label{sec:compare_behaviour}
This technique concerns behavior comparison between VE and \ActE{},
with the aim to check differences. Thirteen interviewees discussed this informal validation technique%
. This is done in different ways at different abstraction levels. %
From the interviews, we observed that one way is using observational tests at high level, where the behavior of \ActE{} and VE are observed together using 3D visualization and checked for synchronicity and differences. At times, specific inputs or measurements from \ActE{}s are given to both \ActE{} and VE and behavior matching is checked. 
We further observed that in cases of DTs used for predictive maintenance, validation is done by initially observing the VE-based predictions and observing and comparing the output of the \ActE{} to those later on to check the accuracy of the predictions.

Moreover, we identified that deeper observational tests are done by creating visual representations such as graphs or 3D visuals of the behavior of both \ActE{} and VE and superimposing them to observe the extent of overlapping and differences. In other cases%
, \ActE{} and VE behaviors are translated into events and actions in a Gantt chart~\cite{wilson2003gantt}.
The timing and sequence of actions are compared between \ActE{} and VE to check if there are any differences. 
When there is a combination of continuous and discrete behaviors,
continuous signals are transformed into discrete ones and then the behaviors of \ActE{} and VE are compared to check for equivalence. %

There are some challenges with this type of validation. Being %
dependent on measured data from the \ActE{}, it is unreliable according to two 
interviewees, due to
incorrect data stemming from measurement errors, faulty equipment, or incorrect interpretations. Moreover, we speculate that there could be other issues in a DT such as consistency issues at runtime which may not be discovered by the aforementioned methods. %

\item \textbf{Formal methods and tools:}
We identified two interviewees from industry and one interviewee from academia who mentioned using formal methods in DT development. Formal verification techniques have been used to validate the behavior of DTs with the help of tools such as Verum's Dezyne~\cite{van2021dezyne} and Coco\footnote{\url{https://cocotec.io/}}%
. %
An interviewee from industry mentioned that DSLs have been used to specify the behavior of a system and transform such specifications into timed automata models in UPPAAL~\cite{behrmann2004tutorial}, allowing model checking. However, model checking is not always scalable enough, considering state space explosion~\cite{baier2008principles}. The interviewee discussed that in order to address this challenge, %
recurring behavioral patterns were identified and validated using model checking, rather than the entire system. In this way, some level of correctness guarantees was %
provided.  %
Another interviewee mentioned that formal methods have been used for consistent execution of models in DTs: formal semantics for such execution were defined and were helpful to understand differences in execution between models. He also shared that
for their model interfaces, they formally proved that the components adhere to the interfaces to avoid interface violations caused by component changes. 
He mentioned that this ensured consistent behavior when integrating components. 
As witnessed by the above, three interviewees mentioned formal techniques for validating DTs; no others did. We speculate that the lower adoption of this method could be attributed to scalability issues.

As mentioned before, formal techniques such as model checking are not scalable, due to state space explosion problems~\cite{baier2008principles}. In addition, we believe that the complexity of formal methods and engineers’ lack of background in them, could possibly contribute to the lower adoption of these techniques.

}
\end{itemize}

\paragraph{Testing and corresponding tools}
Apart from these two categories of techniques identified and quantitatively analysed by us, 
DTs also often undergo testing across their entire lifecycle, in order to check adherence to requirements.
Interviewees used the term ‘testing' to discuss two different items, namely, using scripts to test the system %
and comparing the \ActE{} and VE behavior using observational tests. %
Due to this lack of clarity, 
we did not perform quantitative analysis of interviewee responses for testing.
As mentioned earlier, in some domains such as the space domain, testing is highly time and resource intensive. In such cases, careful consideration is needed on when and what aspects to test, based on the DT and its context.
Moreover, in these cases, testing effort is then balanced with effectiveness. For instance, after resolving an integration issue, only local tests are done. On the other hand, full regression testing would be executed when replacing an entire sub-component. 
An observation worth mentioning here is that this is not only specific to testing DTs, but generally used in the context of testing software systems. %
Different types of tests have been mentioned by interviewees such as model-based testing, integration testing, and unit testing. In some cases, static analysis is also performed to detect coding errors, 
thus, helping gain confidence about the system.
Some tools mentioned by interviewees are Axini's\footnote{\url{https://www.axini.com/en/}}%
, Matlab~\cite{higham2016matlab}, and Unity~\cite{Unity_ref} %
which are used for creating and testing models on the fly. %

\subsubsection{Challenges in validation of DTs} \label{sec:challenges_validation}
Some of the general validation challenges of DTs are presented here.
While quite a few of these challenges were not mentioned by three or more interviewees, we still feel they were relevant to discuss here. Two industrial interviewees expressed that validating a DT is much harder than validating a real system and it is infeasible to validate every aspect of a DT due to its numerous 
degrees of freedom and myriad of parameters. Our study suggests that in some domains, such as the space domain, testing is more resource- and time-intensive than development. An academic interviewee speculated validating a DT to be challenging because the composition of multiple models and the resulting emergent behavior complicate matters. %
Furthermore, interviewees discussed challenges with validating multi-physics models owing to their complexity and thus, not being able to allow real-time
execution. 

Apart from the challenges discussed in the interviews, we identified that the continuous evolution of VEs pose an additional challenge for validation of VEs. This is because of its continuous synchronization with AE, and others as discussed by Zhang et al.~\cite{ZHANG2021151}, which requires specific techniques for continuous validation of VE as it keeps evolving~\cite{Models_meet_data}.
\subsubsection{Strategies for DT validation and to facilitate validation}\label{sec:other_validation}
We list below the strategies which interviewees discussed for validating DTs and for facilitating such validation.
\begin{itemize}
{

\item \textbf{Validation by %
increasing complexity:}
We found three interviewees who advocated an informal bottom up DT validation, gradually increasing complexity. Even when comparing the behavior between \ActE{} and VE, it could be started with simple experiments, followed by more complex ones. The models in a DT could be validated initially and then the integration of models could be validated.
A modular approach can also be adopted where instead of validating the entire DT at once, critical parts of the system are validated initially, followed by other parts and then the integration of all parts.

\item \textbf{Continuous validation of DTs:}
Three interviewees addressed 
continuous validation of DTs. They mentioned that models in a DT undergo updates due to data continuously being communicated from the \ActE{} in the field, feedback from subject matter experts or field service engineers, or bug fixes. When such updates occur, tests are run continuously to validate the DT and ensure that the same overall behavior is exhibited by DT before and after updating.

}
\end{itemize}

\subsubsection{Discussion}\label{sec:discussion_validation}

All but one interviewee currently perform some form of validation of the complete DT or parts of it. In fact, we identified cases where it is necessary to validate the DT continuously as it undergoes changes across its entire lifecycle. %
From this we can infer that validation of DTs is highly important. %
Furthermore, we also identified several challenges involved in validating a DT: one major challenge is that most validation techniques can only cover certain aspects of a DT.
Thus, our study suggests a multi-faceted approach, combining multiple techniques, is required to validate the different aspects of a DT.

\takeaway{We identified 13 out of 19 interviewees who are currently validating their DT by comparing the behavior of \ActE{} and VE. In addition, we found three interviewees who are currently using formal methods to validate their DTs. Moreover, testing has been used as a technique for validating DTs. %
Our analysis suggests that the choice of validation method depends on the DT's purpose, domain, and application; and requires a multi-faceted approach, possibly combining multiple of the aforementioned techniques.
}

\subsection{Properties for validation (RQ7)} \label{sec:validation-properties}

The goal of this research question is to understand which validation properties are considered important and need to be validated in the context of DTs. During the interviews, ten interviewees explicitly mentioned one or more such properties in relation to their respective DTs. We discuss these properties and associated challenges based on the interview data analysis.

\subsubsection{Properties for validation of DTs}\label{sec:intro_properties}
We intended to understand which DT aspects the properties for validation should cover. We believe the properties for validation of DTs could be quantitative or qualitative in nature. 
Two interviewees provided a high level generic overview on this: one industrial interviewee mentioned that the properties to be validated in a DT lie on many levels; another interviewee expressed that the properties should enable the observation of critical things which might go wrong in DTs. 

Interviewees used the terms ‘properties’ and ‘parameters’ while discussing this topic. We identified two important properties, namely, fidelity of VE and time difference in execution between AE and VE. Furthermore, we identified two main categories of properties, namely, properties related to DT's purpose, application, or domain; and properties related to dynamic consistency.
These properties are discussed in detail below 
and are summarised in Table~\ref{tab:V&V_properties}. Moreover, interviewees discussed the different challenges with properties relevant to verification and validation of DTs; we discuss this in Section~\ref{sec:challenges_properties}.

\begin{table}[ht]
    \centering
    \caption{Properties for validation of DTs discussed by industrial (\#I) and academic (\#A) interviewees.}
    \label{tab:V&V_properties}
    \begin{tabular}{p{.6\linewidth}|cc}
        \hline
        \textbf{Properties for validation of DTs} & \textbf{\#I} &\textbf{\#A}\\ \hline
        Fidelity of VE & 8 & 5 \\
        Time difference in execution between AE and VE & 2 & 3 \\
        Related to DT's purpose, application, or domain & 9 & 6 \\
        Related to dynamic consistency & 5 & 2 \\ \hline
    \end{tabular}
\end{table}

\begin{itemize}
{
\item \textbf{Fidelity of VE:} %
From the validation method based on comparing \ActE{} and VE behavior---discussed in Section~\ref{sec:compare_behaviour}---it can be understood that VE fidelity is an important property that interviewees consider. Eight interviewees from industry and five interviewees from academia discussed fidelity of VE as an important property for validation.

\item \textbf{Time difference in execution between \ActE{} and VE:}
While discussing comparing \ActE{} and VE behavior, interviewees also discussed comparing the time difference in execution between \ActE{} and VE. Two interviewees from industry and three interviewees from academia explicitly mentioned that time difference in execution between \ActE{} and VE is an additional property for which validation is needed.
\item \textbf{Related to DT’s purpose, application, or domain:}
We identified six interviewees from academia and nine from industry who conveyed explicitly or implicitly that the properties to validate depend on DT’s purpose, application, or domain. Out of these 15 interviewees, five interviewees from industry and three academics explicitly specified properties which are specific to DT’s purpose, application or domain. Some of these were behavioral properties specific to the DT’s application such as whether the DT behaves correctly during specific scenarios and whether the DT has sufficient information from different sources for decision making. Several temporal properties related to DT’s purpose, application, or domain were also explicitly specified by four interviewees from industry and three from academia. Some of these properties were related to the timing requirements of the domain; specific timeliness properties for certain applications; and real time properties such as software deadlines and activation time.

\item \textbf{Related to dynamic consistency:}
We identified five interviewees from industry and two academic ones who explicitly specified properties which are related to dynamic consistency of DTs. Some of the functional properties to be validated in DTs which were discussed are deadlocks and bottlenecks. %
Several temporal properties were also discussed by three interviewees, related to dynamic consistency issues, such as latency in communication between modeling tools; round-trip time and properties on how swiftly a tool sends and receives messages, and response times. 
}
\end{itemize}

\subsubsection{Challenges with properties for validation of DTs}\label{sec:challenges_properties}
During the interviews, the challenges with validating properties in DTs were discussed. A challenge mentioned by two interviewees from academia was how to measure the quality of DTs and which properties could be used for this. They further mentioned the challenge of quantifying the properties which could be used as a measure of a DT’s quality. This challenge also entailed how these properties could be defined in order for them to be computable. We intended to identify such properties, which need to be validated in a DT.

\subsubsection{Discussion}\label{sec:discussion_validation_properties}
From our findings, we noticed that some interviewees expressed their concerns 
on how to measure the quality of a DT and how to quantify the relevant properties. We also observed that works such as Dalibor et al.~\cite{Cross-Domain_SystMapping} discuss this concern on quality assurance and requirements for DTs. %
Our interviewees, from a range of domains, uniformly agreed that the properties for DT validation depend on the DT's domain, purpose, and application.

\takeaway{Fifteen interviewees discussed that DT properties to be validated depend on a DT's domain, purpose, and application. 
In addition, we identified thirteen interviewees agreeing to fidelity of VE; and five interviewees agreeing to time difference in execution between \ActE{} and VE as important properties for validation of DTs.
Specific functional and temporal properties are of interest to seven interviewees as being key to address dynamic consistency issues in a DT.
    }

\section{Additional findings} \label{sec:additional_findings_ext}
As mentioned in Section~\ref{sec:methodology}, this research was conducted using semi-structured interviews, allowing the interviewee to discuss any topic. In this section we present results 
that do not directly fit any of our research questions. %
Technical topics
mentioned by at least seven interviewees are discussed below, and summarized in Table~\ref{tab:additional_findings}.

Furthermore, given the nature of Digital Twins, the role of humans in DTs is discussed briefly in Section~\ref{sec:Human_role}. Finally, Section~\ref{sec:future_vision} briefly discusses future visions for DTs based on what the interviewees expressed.

\begin{table}[h!]
\caption{Overview of additional findings discussed by industrial (\#I) and academic (\#A) interviewees.  %
}
\label{tab:additional_findings}
\centering
\begin{tabular}{c|cc}
\hline
\multicolumn{1}{c|}{\textbf{Topic}}  & \textbf{\#I} & \textbf{\#A} \\ \hline
Architecture             & 7   & 4   \\
Process                  & 3   & 4   \\
Role of goals in design    & 6   & 5   \\ \hline
\end{tabular}
\end{table}

\subsection{Architecture}

This section aims to explain the DT’s architectural choices as shared by 11 interviewees. The main objective of the architecture is to aid rapid DT development, for which two key architectural properties were identified: re-usability of components and maintainability of the DT. We observed four architectures mentioned by the interviewees, but we only report on the one with at least three mentions, which is the block-based architecture mentioned by four industrial and two academic interviewees. For each interviewee the entity encapsulated in such a block is different. For two academic interviewees, the block is a model which must be configurable, that can exist at different levels of abstraction, e.g., a component of a machine, the complete machine, or the entire manufacturing system. Another two industrial interviewees explained that in order to facilitate VE maintainability, their notion of a block is a component of a machine or process, but never the entire system.
According to our analysis, block-based architectures with their separation of concerns between entities of a different nature—such as components of the AE or machines cluster—aid rapid DT development, due to component re-usability and maintainability.

\subsection{Process}

Seven interviewees discussed the process to build a DT. Five of them follow a software development process adjusted to DT development, although they did not specify how exactly this adjustment was made. Another two described specific, domain-dependent, processes to design a DT. The software development processes that were mentioned are DevOps (Software Development and Operations) or Agile, which facilitate cross-domain cooperation. The other two interviewees used specific design processes and software development practices in different stages of their DT development. While the academic interviewee shared that he uses practices from the V-model to perform unit testing, the industrial interviewee discussed that he uses Continuous Integration-Continuous Deployment (CI/CD) tools.

\subsection{Role of goals in design}

This section discusses the role that a clear goal can play in the design of digital twins, according to 11 interviewees. All of them mentioned determining a goal (or purpose, or service) as the first step to developing a DT. We identified the influence of the goal in three entities, namely, the model, data, and other design choices such as tool selection and resource definition. Related to models, they mentioned that the goal defines the model’s fidelity, level of abstraction, type (e.g., continuous time), and modeling approach (e.g., data or physics-based). Concerning data, the goal defines the data to collect, data processing methods, and the selection of sensors and actuators.

\subsection{Role of humans in DTs} \label{sec:Human_role}

\begin{figure}[h!]
    \centering

    \includegraphics[width=\linewidth]{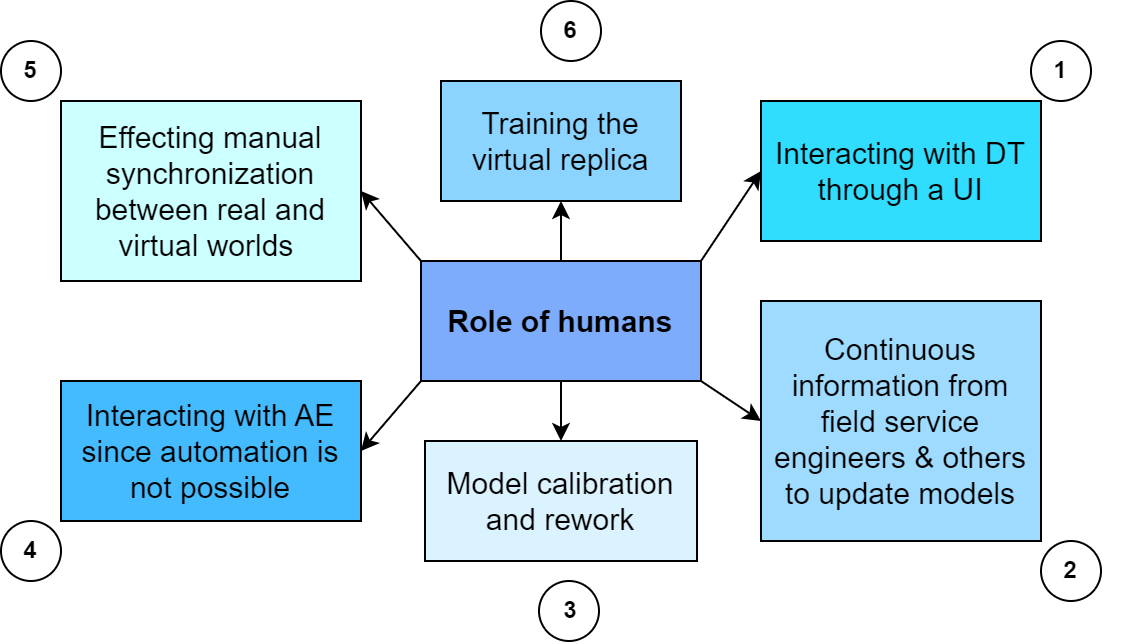}
    \caption{Role of humans in the operation of a DT.}
    \label{fig:Role_of_humans}
\end{figure}

We identified two interviewees who explicitly mentioned that humans play an important role in a DT and should also be seen as a part of a DT. Several interviewees also implicitly discussed the importance of humans in a DT during its operation and this has been visualised in Figure~\ref{fig:Role_of_humans}, with the numbers used to guide the reader in the text below. One common observation from our analysis is the role of humans in \Circled{1} for monitoring, training, and other purposes. Three interviewees from academia also mentioned that continuous information input and knowledge use from humans as shown in \Circled{2} help in updating and improving the models continuously while the DT is in operation. One industrial interviewee also mentioned that \Circled{3} which involves changing and tuning parameters in the model based on the data from the real world, may require human intervention. Our analysis further suggests that certain services provided by the DT cannot be automated and thus, a human is required to translate the information received through the user interface, into actions on the AE  (see \Circled{4}) and this was explicitly mentioned by two interviewees. Three mentioned that in some of the DTs they have worked on, there is no direct connection between the AE and its virtual counterpart and thus, humans act as the bridge connecting these two entities by manually transferring data from the AE to its corresponding VE for \Circled{5}. Apart from these, one interviewee from academia mentioned that in one of the DTs he had worked with, humans played an important role in training \Circled{6} the AE for its specific purpose. Thus, it has been observed that humans, apart from contributing to the development of DTs, play an important role in the operation of DTs.

\subsection{Future vision on DTs}
\label{sec:future_vision}
One interview question was aimed to understand interviewees’ perspective on the future evolution of DTs. During our analysis, we identified interesting and significant outcomes from this discussion. Two interviewees from industry expressed that the DTs could evolve in the future to become a tradeable asset which would be made available alongside its corresponding AE: when an AE is traded between two parties, its corresponding DT could also be traded or provided access to. Moreover, two interviewees shared that in the future, DTs from different domains around us could probably then interact with each other and exchange useful information for use-cases such as improving decision making, accurate diagnosis, etc. According to one interviewee from academia, machine learning and reinforcement learning could possibly be combined with DTs in the future, to help to learn about complex systems (i.e., safety critical systems) in a virtual environment, when this is difficult to do on the real-world system. In addition, two interviewees emphasized that in the future, AEs could possibly become more autonomous and self-adaptive with the help of their DTs. Furthermore, two interviewees predicted DT usage to enable increased automation of real world processes and improving design support. Additionally, we identified two interviewees who predicted that there would be a shift in the way AEs are developed in the future, where a DT would always be completely part of the development of systems.

We also found interviewees discussing various visions on future improved DT development. Two industrial interviewees mentioned that they believe that standards would become available for, e.g., developing and maintaining DTs, and for managing and combining data in DTs. In addition, two interviewees predicted improvements in the ease-of-use and intuitiveness of DT development tools which would enable experts from different domains with minimal software knowledge and expertise to develop DTs with ease. Moreover, two interviewees explicitly mentioned that DTs should evolve over their entire lifecycle to serve new purposes, i.e., provide new services in addition to what they were initially developed for. Furthermore, one interviewee from industry mentioned that visualization in DTs would possibly improve in the future which would pave a way for better interpretations.

\subsection{Discussion on additional findings}
During the interview, the interviewees shared their thoughts and opinions on the development of DTs. The additional findings, as visualised in Table~\ref{tab:additional_findings}, are related to architecture, development process and the role of goals in DT design. In addition, the role of humans in DTs and the future vision on DTs are discussed.

Most interviewees who discussed DT architecture mentioned a block-based architecture as their preference. Each has a different interpretation of what a block entails, 
but all agree that the aim of this architecture is rapid DT development and maintenance. 

From the interviewees who discussed their process for DT development, the majority stated using an adapted software development process, such as Dev-Ops or Agile. Others who use specific development processes, seem to use some software development practices and tools, such as unit testing or CI/CD tools. Our analysis shows that the main driver to use software practices or processes is the software-centered nature of DT's. %

The eleven interviewees %
who discussed the role of the goal in the design, agreed that it is crucial to define it before starting DT development. The goal has a big influence on many design decisions related to models, data, and tooling. Examples of this are the fidelity of the models or the data selection from the \ActE{}.

Furthermore, while several interviewees implicitly discussed the importance of humans' role in DTs, only two interviewees explicitly considered humans to be a part of DTs.

Future predictions were made on three different aspects of DTs, namely, the trade aspect of DTs, the influence of DTs in engineering applications, and the evolution of DT development. 
We believe that some of the aforementioned future applications of DTs such as increasing AE’s autonomy and adaptability, and improving automation and design support, are highly important. We further believe that the influence of DTs in engineering applications could substantially grow in the future.

\vspace{1em}
\noindent\fbox{
    \parbox{.96\linewidth}{
        \textbf{Takeaway message:}\\
        Our analysis shows that DT architectures should facilitate maintainability, cross-domain collaboration, and rapid development. 
        It also shows that DT development processes contain ingredients of software development processes or adaptations thereof for DTs. A common recommendation for development is to define the goal of the DT before starting its development. Predictions on the business model of DTs have been made such as DTs and their related data becoming a tradeable asset whose ownership could be transferred or shared across the lifecycle. DTs have been predicted to improve the automation and self-adaptability of systems; and also to help in the design support for such systems. Current challenges in DT development such as lack of intuitiveness and standards are predicted to be resolved in the future. }
}

\section{Threats to validity}
\label{sec:threats}
This research is an empirical study, which is never completely devoid of omissions or pitfalls. We discuss the threats to validity (TTV) as described by Easterbrook et al.~\cite{easterbrook2008selecting} and the measures taken to mitigate these.

\textbf{Construct validity} pertains to quality of the measurement of the constructs for the experiment. One concern regarding this TTV is the level of expressiveness of information shared by the interviewees about the DT practices and technical DT issues at their organization. However, this was mitigated to a certain extent by emphasizing the research's approval by the universities' ethics review board and data stewards, assuring data privacy rights and complete anonymization of the collected data and its processing.

A second TTV concerns the \textbf{limited variability of interviewee opinions}. To mitigate, we included interviewees from both industry and academia with varied educational backgrounds, different levels of experience with DTs; working in diverse roles; from different industrial domains; and from companies of different sizes. As discussed in Section~\ref{sec:find_interviewee}, most of the interviewees are from our own network comprising experts from start-ups to high-tech manufacturing companies from the `Brainport region', which is one of Europe’s biggest industrial hubs. However, we made sure to include interviewees from different backgrounds such as manufacturing and chemical processing, high-tech products, building and construction, transport and logistics, information systems and healthcare.
In addition, we also included interviewees working on diverse applications of DTs, namely control, design and development, predictive maintenance, monitoring, analysis, testing, training and demonstration. 

A third TTV concerns the \textbf{understandability of the interview questions}. To mitigate this threat, we conducted a pilot interview with an interviewee from industry working with DTs.
Based on the feedback received, interview questions were reordered and rephrased for a seamless interview flow and ease of understanding, the process of which is explained in Section~\ref{sec:question_design}.

The fourth TTV is related to the \textbf{background of the interviewers} which may affect the outcome of interviews. This was mitigated by not having one interviewer but three. Moreover, these interviewers had diverse backgrounds in different domains such as software engineering, embedded systems, chemical, automotive, mathematics, electronics, and instrumentation engineering. In addition to diverse backgrounds, the interviewers had experience working in both industry and academia. 

A last construct TTV is related to \textbf{interviewer bias}. This was mitigated by creating an interview guide which had instructions to drive the interview with an exhaustive list of pre-defined questions to be asked. Furthermore, each interview was conducted by two interviewers, one asking questions and one keeping track of questions and making notes. %

Regarding \textbf{internal validity}, the dependency of results on just the interview data is a potential threat. In order to mitigate this, the data analysis was done meticulously following the analysis methodology described in Section~\ref{sub:data-analysis}. Data analysis involved three levels of transcription: firstly, automated transcription; secondly, manual transcription by student assistants; and thirdly, verification of the entire transcripts against the recorded interviews by three researchers. The coding process was scrupulously done by three researchers in an interpretative manner over two months. For any artifact, the process was deemed complete only when it was coded equally by at least two researchers. When the two researchers had different interpretations in coding a particular artifact, these were presented and discussed among the researchers to arrive at a common understanding on such codes. This way of working was described in Section~\ref{sec:label_and_topic_generation}. 

Regarding \textbf{external validity}, the generalizability of results is a TTV. We tried to minimize this TTV by having a significant number of interviewees with diverse roles and varied educational backgrounds from both industry and academia; with different levels of experience with DTs; from different industrial domains varying from health care to space exploration; from companies of different sizes; and with different levels of understanding of DTs, considering the lack of common understanding of DTs in both industry and academia.
However, the sample size of nineteen in this research can still be seen as relatively small, possibly limiting external validity.

\section{Related work } \label{sec:discussion} 

In this section, we explore the academic literature relevant to our study, focusing on current practices and challenges in the implementation of DTs. \cref{tab:RW_papers} contains a summary of the relevant papers we have explored and compared to our current research. 

\begin{table*}
\centering
\caption{Summary of related work research related to the implementation of DTs. SLR = Systematic Literature Review.}
\label{tab:RW_papers}
\begin{tabular}{ l | p{.55\linewidth} |l}
 \hline
\textbf{Reference} & \textbf{Summary} & \textbf{Approach} \\ \hline
Perno et~al.~\cite{RW_ProcessInds} & This paper conducts a study to identify categories of barriers hindering the development of DTs in the process industry. The paper proposes a model designed to recognize potential facilitators addressing the identified barriers and intends to offer possible solutions. & SLR \\ \hline
Segovia et~al.~\cite{RW_Model_imp} & This paper analyses the state-of-practice on DT development, with a specific focus on the methodological approaches found in scientific literature. The paper considers three primary phases: design, modeling, and implementation.  It states the challenges that are encountered in each phase. & SLR \\ \hline
Sharma et~al.~\cite{DT_state_art_1} & This study focuses on the conceptualization and adoption of DT in practical applications. The goal of this research is to define the key components of DTs and identify open research questions for their development. This study analyses three case studies; two from the construction domain and one from the medical domain. & SLR \\ \hline
Liu et~al.~\cite{RW_SoA_survey} & This paper performs an analysis of the approaches to develop DT based on a classification. Specifically, its implementation classification criteria are based on a central component. For instance, if the primary component is a service, the DT is categorized as a "DT as a   service.   & SLR \\ \hline
Trauer et~al.~\cite{RW_ImpChallenge} & The objective of this paper is to investigate the obstacles faced by 61 practitioners in implementing DTs.   It particularly aims to understand the characteristics of DTs and the technical and non-technical challenges associated with their implementation. & Interview \\ \hline
Mihai et~al.~\cite{9899718} & This research aims to understand the technologies that ease the implementation of DT and its challenges. It identifies enabling technologies and outlines research challenges for DT   implementation, including fidelity and synchronization of models and ongoing standardization efforts.  & SLR \\ \hline
Botin-Sanabria et~al.~\cite{RW_challengeApp} & This paper examines different barriers to the use and development of DTs, revealing high-level obstacles spanning social, technical, and business domains. & SLR \\ \hline
Parnianifard et~al.~\cite{RW_survey_DTandCPS} & This research discusses limitations and future challenges, highlighting the complexity and interoperability issues arising from the heterogeneous nature of DT   components. & SLR \\ \hline
Ammar et~al.~\cite{RW_DT_construction} & This study aims to understand the usage and challenges of DTs by interviewing eight practitioners in the construction industry. Six categories of challenges are identified, with a   focus on technology development, particularly in sensor technology. & Interview \\ \hline
\end{tabular}
\end{table*}

We have identified two main research topics from the papers presented in \cref{tab:RW_papers}. The first domain is associated with methodologies employed in the implementation DTs~\cite{RW_Model_imp,DT_state_art_1,RW_SoA_survey,RW_DT_construction}. Research related with methodologies for implementing DTs has yielded valuable outputs by identifying potential open research questions and approaches embraced by practitioners. However, it is noteworthy that the focus of these investigations predominantly revolves around the primary approaches by practitioners. This differs from our research where we focus our effort on specific development areas, such as cross-domain model consistency, orchestration, and dynamic consistency.

The second topic revolves around the identification of challenges and enablers in the implementation of DTs~\cite{RW_ProcessInds,RW_ImpChallenge,9899718,RW_challengeApp,RW_survey_DTandCPS}. Each of these works has highlighted diverse technical and non-technical challenges confronted by practitioners in the course of DT implementation. Nevertheless, these studies avoid examining deeply the strategies employed by practitioners in addressing these challenges. Our research focused particularly on exploring the strategies used by practitioners to deal with different implementation challenges. 

\section{Conclusion} \label{sec:conclusion}

In this exploratory research, we studied the current landscape of DTs from a technical point of view, particularly its current state-of-practice on design, development, operation and maintenance. To do that we interviewed individuals from industry and academia. Our research focused on software aspects related to DTs, specifically the interviewees' understanding of DTs, model consistency, integration, orchestration and validation. In addition, we also discussed our interviewees' opinions on the future impact of DTs. 

Our research aims to understand the diverse practices, methods, and tools employed by professionals engaged in the design, development, and maintenance of DTs. Our specific research focuses on discerning the methods employed to ensure cross-domain consistency, orchestrating DT applications, and maintaining dynamic consistency.

The empirical findings show a lack of methods and tools designed to address inconsistencies between models. Furthermore, we found that distinct tools and approaches are used for the integration and orchestration of heterogeneous models, with their selection dependent on the specific domain or nature of the DT application. Regarding dynamic consistency, approaches primarily involved the comparative analysis between virtual and physical entities.

Our research indicates research opportunities. (1) Developing methods to deal with consistency, and formulating guidelines for reuse components in DTs; (2) developing tools that facilitate integration and orchestration; and (3) determining the required properties and approaches for validating dynamic consistency of DT.

Our analyses of the interviews showed many challenges around DT development, maintenance, and operations. Yet also emerging were various ways to potentially address these challenges, involving concepts and results from more classical software and systems engineering. The field holds a lot of promise for, requires applying and adapting such concepts and results. This research on how best to adjust and apply these to DT development, maintenance, and operations. With such research, and with the vision of the future as expressed by the interviewees, the future definitely looks both bright and twinned.

\section*{Acknowledgement}
This research was funded by NWO (the Dutch national research council) under the NWO AES Perspectief program, project code P18-03 P3.

\printbibliography

@String{Computer = "{IEEE} Computer" }

@String{Springer = "Springer-Verlag" }

@article{lama2023, 
    doi = {10.21105/joss.05135}, 
    year = {2023}, 
    publisher = {The Open Journal}, 
    volume = {8}, 
    number = {85}, 
    pages = {5135}, 
    author = {Victoria Bogachenkova and Eduardo Costa Martins and Jarl Jansen and Ana-Maria Olteniceanu and Bartjan Henkemans and Chinno Lavin and Linh Nguyen and Thea Bradley and Veerle Fürst and Hossain Muhammad Muctadir and Mark van den Brand and Loek Cleophas and Alexander Serebrenik}, 
    title = {LaMa: a thematic labelling web application}, 
    journal = {Journal of Open Source Software} 
}

@Inbook{magaldi2020semi,
    author="Magaldi, Danielle
    and Berler, Matthew",
    editor="Zeigler-Hill, Virgil
    and Shackelford, Todd K.",
    title="Semi-structured Interviews",
    bookTitle="Encyclopedia of Personality and Individual Differences",
    year="2020",
    publisher="Springer International Publishing",
    address="Cham",
    pages="4825--4830",
    isbn="978-3-319-24612-3",
    doi="10.1007/978-3-319-24612-3_857",
}

@article{ZHANG2021151,
    title = {Building a right digital twin with model engineering},
    journal = {Journal of Manufacturing Systems},
    volume = {59},
    pages = {151-164},
    year = {2021},
    issn = {0278-6125},
    doi = {10.1016/j.jmsy.2021.02.009},
    author = {Lin Zhang and Longfei Zhou and Berthold K.P. Horn}
}

@article{Strandberg2019EthicalEngineering,
    title = {{Ethical Interviews in Software Engineering}},
    year = {2019},
    journal = {International Symposium on Empirical Software Engineering and Measurement},
    author = {Strandberg, Per Erik},
    month = {9},
    volume = {2019-September},
    publisher = {IEEE Computer Society},
    isbn = {9781728129686},
    doi = {10.1109/ESEM.2019.8870192},
    issn = {19493789},
    pages={1-11},
    arxivId = {1906.07993}
}

@article{Kiger2020,
   author = {Michelle E. Kiger and Lara Varpio},
   issn = {0142-159X},
   issue = {8},
   journal = {Medical Teacher},
   month = {8},
   pages = {846-854},
   publisher = {Taylor and Francis Ltd},
   title = {Thematic analysis of qualitative data: AMEE Guide No. 131},
   volume = {42},
   year = {2020},
   doi={10.1080/0142159X.2020.1755030}
}

@article{castillo2016preparing,
    title={Preparing for interview research: The interview protocol refinement framework},
    author={Castillo-Montoya, Milagros},
    journal={The qualitative report},
    volume={21},
    number={5},
    pages={811--831},
    year={2016},
    publisher={Nova Southeastern University, Inc.},
    doi={10.46743/2160-3715/2016.2337}
}

@INPROCEEDINGS{Hove2005ExperiencesResearch,  
    author={Hove, S.E. and Anda, B.},  
    booktitle={11th IEEE International Software Metrics Symposium (METRICS'05)},
    title={Experiences from conducting semi-structured interviews in empirical software engineering research},   
    year={2005},  
    volume={},  
    number={},  
    pages={10 pp.-23},  
    doi={10.1109/METRICS.2005.24}}

@article{Mohagheghi2007,
  doi = {10.1007/s10664-007-9040-x},
  year = {2007},
  month = may,
  publisher = {Springer Science and Business Media {LLC}},
  volume = {12},
  number = {5},
  pages = {471--516},
  author = {Parastoo Mohagheghi and Reidar Conradi},
  title = {Quality,  productivity and economic benefits of software reuse: a review of industrial studies},
  journal = {Empirical Software Engineering}
}

@ARTICLE{Tao_DT-in-Industry,
  author={Tao, Fei and Zhang, He and Liu, Ang and Nee, A. Y. C.},
  journal={IEEE Transactions on Industrial Informatics}, 
  title={Digital Twin in Industry: State-of-the-Art}, 
  year={2019},
  volume={15},
  number={4},
  pages={2405-2415},
  doi={10.1109/TII.2018.2873186}}

@InProceedings{baker2005,
    author="Baker, Paul
    and Loh, Shiou
    and Weil, Frank",
    editor="Briand, Lionel
    and Williams, Clay",
    title="Model-Driven Engineering in a Large Industrial Context --- Motorola Case Study",
    booktitle="Model Driven Engineering Languages and Systems",
    year="2005",
    publisher="Springer Berlin Heidelberg",
    address="Berlin, Heidelberg",
    pages="476--491",
    isbn="978-3-540-32057-9",
    doi = {10.1007/11557432_36}
}

@article{DAVIES201488,
    title = {Model-driven engineering of information systems: 10 years and 1000 versions},
    journal = {Science of Computer Programming},
    volume = {89},
    pages = {88-104},
    year = {2014},
    note = {Special issue on Success Stories in Model Driven Engineering},
    issn = {0167-6423},
    doi = {10.1016/j.scico.2013.02.002},
    author = {Jim Davies and Jeremy Gibbons and James Welch and Edward Crichton},
}

@article{RODRIGUESDASILVA2015139,
    title = {Model-driven engineering: A survey supported by the unified conceptual model},
    journal = {Computer Languages, Systems \& Structures},
    volume = {43},
    pages = {139-155},
    year = {2015},
    issn = {1477-8424},
    doi = {10.1016/j.cl.2015.06.001},
    author = {Alberto {Rodrigues da Silva}},
}

@article{DEARAUJOSILVA2021101021,
    title = {A survey of Model Driven Engineering in robotics},
    journal = {Journal of Computer Languages},
    volume = {62},
    pages = {101021},
    year = {2021},
    issn = {2590-1184},
    doi = {10.1016/j.cola.2020.101021},
    author = {Edson {de Araújo Silva} and Eduardo Valentin and Jose Reginaldo Hughes Carvalho and Raimundo {da Silva Barreto}},
}

@article{Wright2020,
   author = {Louise Wright and Stuart Davidson},
   doi = {10.1186/s40323-020-00147-4},
   issn = {2213-7467},
   issue = {1},
   journal = {Advanced Modeling and Simulation in Engineering Sciences},
   month = {12},
   pages = {13},
   title = {How to tell the difference between a model and a digital twin},
   volume = {7},
   year = {2020},
}

@inbook{behrmann2004tutorial,
    author="Behrmann, Gerd
    and David, Alexandre
    and Larsen, Kim G.",
    editor="Bernardo, Marco
    and Corradini, Flavio",
    title="A Tutorial on Uppaal",
    bookTitle="Formal Methods for the Design of Real-Time Systems: International School on Formal Methods for the Design of Computer, Communication, and Software Systems, Bertinora, Italy, September 13-18, 2004, Revised Lectures",
    year="2004",
    publisher="Springer Berlin Heidelberg",
    address="Berlin, Heidelberg",
    pages="200--236",
    isbn="978-3-540-30080-9",
    doi="10.1007/978-3-540-30080-9_7",
}

@article{DT_state_art_1,
	author = {Angira Sharma and Edward Kosasih and Jie Zhang and Alexandra Brintrup and Anisoara Calinescu},
	doi = {10.1016/j.jii.2022.100383},
	issn = {2452-414X},
	journal = {Journal of Industrial Information Integration},
	pages = {100383},
	title = {Digital Twins: State of the art theory and practice, challenges, and open research questions},
	volume = {30},
	year = {2022}
}

@inproceedings{walravensTurtle,
author={Walravens, Gijs and Muctadir, Hossain Muhammad and Cleophas, Loek},
booktitle={ACM/IEEE 25th International Conference on Model Driven Engineering Languages and Systems (MODELS '22 Companion)},
title={Virtual Soccer Champions: A Case Study on Artifact Reuse in Soccer Robot Digital Twin Construction},
year={2022},
volume={},
number={},
location={Montreal, QC, Canada},
doi={10.1145/3550356.3561586}
}

@misc{hause2006sysml,
  title={The {SysML} modelling language},
  author={Hause, Matthew and others},
  booktitle={Fifteenth European Systems Engineering Conference},
  howpublished = {\url{https://www.omgsysml.org/The_SysML_Modelling_Language.pdf}},
  note = {Accessed 12 May 2023}
}

@misc{om2017unified,
  title={Unified Modeling Language Specification Version 2.5.1},
  author={{Object Management Group}},
  howpublished = {\url{https://www.omg.org/spec/UML/}},
  note = {Accessed 17 November 2022}
}

@INPROCEEDINGS{Models_meet_data,
  author={van den Brand, Mark and Cleophas, Loek and Gunasekaran, Raghavendran and Haverkort, Boudewijn and Manrique-Negrin, David A.  and Muctadir, Hossain Muhammad},
  booktitle={2021 ACM/IEEE International Conference on Model Driven Engineering Languages and Systems Companion (MODELS-C)}, 
  title={Models Meet Data: Challenges to Create Virtual Entities for Digital Twins}, 
  year={2021},
  volume={},
  number={},
  pages={225-228},
  doi={10.1109/MODELS-C53483.2021.00039}}

@article{BARROSJUSTO20181,
	author = {Jos{\'e} L. Barros-Justo and Fernando Pinciroli and Santiago Matalonga and Nelson Mart{\'\i}nez-Araujo},
	doi = {10.1016/j.infsof.2018.06.003},
	issn = {0950-5849},
	journal = {Information and Software Technology},
	pages = {1-21},
	title = {What software reuse benefits have been transferred to the industry? A systematic mapping study},
	volume = {103},
	year = {2018}
}

@misc{muctadir_2023_10187933,
  author       = {Muctadir, Hossain Muhammad and
                  Manrique Negrin, David A. and
                  Gunasekaran, Raghavendran and
                  Cleophas, Loek and
                  van den Brand, Mark and
                  Haverkort, Boudewijn R.},
  title        = {{Replication package for the interview study on 
                   current trends in Digital Twin development,
                   maintenance, and operation}},
  month        = nov,
  year         = 2023,
  publisher    = {Zenodo},
  doi          = {10.5281/zenodo.10187933}
}

@ARTICLE{9899718,  author={Mihai, Stefan and Yaqoob, Mahnoor and Hung, Dang V. and Davis, William and Towakel, Praveer and Raza, Mohsin and Karamanoglu, Mehmet and Barn, Balbir and Shetve, Dattaprasad and Prasad, Raja V. and Venkataraman, Hrishikesh and Trestian, Ramona and Nguyen, Huan X.},  journal={IEEE Communications Surveys \& Tutorials},   title={Digital Twins: A Survey on Enabling Technologies, Challenges, Trends and Future Prospects},   year={2022},  volume={24},  number={4},  pages={2255-2291},  doi={10.1109/COMST.2022.3208773}}

@article{grieves2017digital,
    title={Digital twin: manufacturing excellence through virtual factory replication},
  author={Grieves, Michael},
  journal={White paper},
  volume={1},
  doi={10.13140/RG.2.2.26367.61609},
  url={https://www.researchgate.net/publication/275211047_Digital_Twin_Manufacturing_Excellence_through_Virtual_Factory_Replication},
  pages={1--7},
  year={2014},
  publisher={Florida Institute of Technology}
}

@article{DT_concept_tech_applications,
title = {Review of digital twin about concepts, technologies, and industrial applications},
journal = {Journal of Manufacturing Systems},
volume = {58},
pages = {346-361},
year = {2021},
note = {Digital Twin towards Smart Manufacturing and Industry 4.0},
issn = {0278-6125},
doi = {10.1016/j.jmsy.2020.06.017},
author = {Mengnan Liu and Shuiliang Fang and Huiyue Dong and Cunzhi Xu}
}

@article{torres2020systematic,
   author = {Weslley Torres and Mark G. J. van den Brand and Alexander Serebrenik},
   doi = {10.1007/s10270-020-00834-1},
   issn = {1619-1366},
   journal = {Software and Systems Modeling},
   month = {10},
   publisher = {Springer},
   title = {A systematic literature review of cross-domain model consistency checking by model management tools},
   year = {2020},
}

@book{baier2008principles,
title = "Principles of Model Checking",
author = "Christal Baier and Katoen, {Joost P.}",
year = "2008",
month = may,
language = "English",
isbn = "978-0-262-02649-9",
publisher = "MIT Press",
address = "United States"}

@article{interviews_DT_Infrastructure,
author = {Didem Gürdür Broo and Jennifer Schooling},
title = {Digital twins in infrastructure: definitions, current practices, challenges and strategies},
journal = {International Journal of Construction Management},
volume = {23},
number = {7},
pages = {1254-1263},
year  = {2023},
publisher = {Taylor & Francis},
doi = {10.1080/15623599.2021.1966980},
}

@article{Hisarciklilar2011,
   author = {Onur Hisarciklilar and Keyvan Rahmani and Vince Thomson},
   doi = {10.1115/DETC2010-28464},
   isbn = {9780791844144},
   journal = {Proceedings of the ASME Design Engineering Technical Conference},
   month = {3},
   pages = {555-563},
   publisher = {American Society of Mechanical Engineers Digital Collection},
   title = {A Conflict Detection Approach for Collaborative Management of Product Interfaces},
   volume = {6},
   year = {2011},
}

@article{Cross-Domain_SystMapping,
title = {A Cross-Domain Systematic Mapping Study on Software Engineering for Digital Twins},
journal = {Journal of Systems and Software},
volume = {193},
pages = {111361},
year = {2022},
issn = {0164-1212},
doi = {10.1016/j.jss.2022.111361},
author = {Manuela Dalibor and Nico Jansen and Bernhard Rumpe and David Schmalzing and Louis Wachtmeister and Manuel Wimmer and Andreas Wortmann}
}

@inproceedings{bielefeldt2015computationally,
    author = {Bielefeldt, Brent and Hochhalter, Jacob and Hartl, Darren},
    title = "{Computationally Efficient Analysis of SMA Sensory Particles Embedded in Complex Aerostructures Using a Substructure Approach}",
    volume = {1: Development and Characterization of Multifunctional Materials; Mechanics and Behavior of Active Materials; Modeling, Simulation and Control of Adaptive Systems},
    series = {Smart Materials, Adaptive Structures and Intelligent Systems},
    year = {2015},
    month = {09},
    doi = {10.1115/SMASIS2015-8975}
}

@article{el2018digital,
  author={El Saddik, Abdulmotaleb},
  journal={IEEE MultiMedia}, 
  title={Digital Twins: The Convergence of Multimedia Technologies}, 
  year={2018},
  volume={25},
  number={2},
  pages={87-92},
  doi={10.1109/MMUL.2018.023121167}
}

@article{van2021dezyne,
    doi = {10.4204/eptcs.338.4},
	year = 2021,
	month = {aug},
	publisher = {Open Publishing Association},
	volume = {338},
	pages = {19--30},
	author = {Rutger van Beusekom and Bert de Jonge and Paul Hoogendijk and Jan Nieuwenhuizen},
	title = {Dezyne: Paving the Way to Practical Formal Software Engineering}, 
	journal = {Electronic Proceedings in Theoretical Computer Science}
}

@book{higham2016matlab,
 author = "Desmond J. Higham and Nicholas J. Higham",
 title = "{MATLAB} Guide",
 edition = "Third",
 publisher = "Society for Industrial and Applied Mathematics",
 address = "Philadelphia, PA, USA",
 year = 2017,
 pages = "xxvi+476",
 isbn = "978-1-61197-465-2"
 
     }

@article{tao2017digital,
    title = {Digital twin driven prognostics and health management for complex equipment},
    journal = {CIRP Annals},
    volume = {67},
    number = {1},
    pages = {169-172},
    year = {2018},
    issn = {0007-8506},
    doi = {10.1016/j.cirp.2018.04.055},
    author = {Fei Tao and Meng Zhang and Yushan Liu and A.Y.C. Nee}
}

@article{FMU_in_DTs,
title = {FMU-supported simulation for CPS Digital Twin},
journal = {Procedia Manufacturing},
volume = {28},
pages = {201-206},
year = {2019},
note = {7th International conference on Changeable, Agile, Reconfigurable and Virtual Production (CARV2018)},
issn = {2351-9789},
doi = {10.1016/j.promfg.2018.12.033},
author = {Elisa Negri and Luca Fumagalli and Chiara Cimino and Marco Macchi}
}

@misc{Ptolemaeus:14:SystemDesign,
  title={System design, modeling, and simulation: using Ptolemy II},
  author={Ptolemaeus, Claudius},
  organization={Ptolemy.org Berkeley},
  howpublished ={\url{https://ptolemy.berkeley.edu/books/Systems/}},
  note = {Accessed 02 January 2023},
}

@misc{IBMUserGuide,
author={{IBM Corporation}},
title = {Rational Rhapsody User Guide},
organization = {IBM Corporation},
adress = {1 New Orchand Rd, Armond, New York 10504, U.S.A.},
howpublished = 
{\url{https://public.dhe.ibm.com/software/rationalsdp/documentation/product_doc/Rhapsody/version_7-5/UserGuide.pdf}},
note = {Accessed 02 January 2023},
}

@misc{Unity_ref,
    author = {{Unity Technologies}},
    title = {Unity User Manual (2019.3)},
    howpublished = {\url{https://docs.unity3d.com/Manual/index.html}},
    note = {Accessed 02 January 2023},
}

@article{wilson2003gantt,
title = {Gantt charts: A centenary appreciation},
journal = {European Journal of Operational Research},
volume = {149},
number = {2},
pages = {430-437},
year = {2003},
note = {Sequencing and Scheduling},
issn = {0377-2217},
doi = {10.1016/S0377-2217(02)00769-5},
author = {James M. Wilson}
}

@InProceedings{Standard_orchestration,
author="Barros, Alistair
and Dumas, Marlon
and Oaks, Phillipa",
editor="Bussler, Christoph J.
and Haller, Armin",
title="Standards for Web Service Choreography and Orchestration: Status and Perspectives",
booktitle="Business Process Management Workshops",
year="2006",
publisher="Springer Berlin Heidelberg",
address="Berlin, Heidelberg",
pages="61--74",
isbn="978-3-540-32596-3",
doi={10.1007/11678564_7}
}

@article{BP_heterogeneousSys,
title = {Business processes oriented heterogeneous systems integration platform for networked enterprises},
journal = {Computers in Industry},
volume = {61},
number = {2},
pages = {127-144},
year = {2010},
note = {Integration and Information in Networked Enterprises},
issn = {0166-3615},
doi = {10.1016/j.compind.2009.10.009},
author = {Qing Li and Jian Zhou and Qi-Rui Peng and Can-Qiang Li and Cheng Wang and Jing Wu and Bei-En Shao},

}

@article{easterbrook2008selecting,
  title={Selecting empirical methods for software engineering research},
  author={Easterbrook, Steve and Singer, Janice and Storey, Margaret-Anne and Damian, Daniela},
  journal={Guide to advanced empirical software engineering},
  pages={285--311},
  year={2008},
  publisher={Springer},
  doi = {http://dx.doi.org/10.1007/978-1-84800-044-5}
}

@article{RW_ProcessInds,
title = {Implementation of digital twins in the process industry: A systematic literature review of enablers and barriers},
journal = {Computers in Industry},
volume = {134},
pages = {103558},
year = {2022},
issn = {0166-3615},
doi = {https://doi.org/10.1016/j.compind.2021.103558},
url = {https://www.sciencedirect.com/science/article/pii/S0166361521001652},
author = {Matteo Perno and Lars Hvam and Anders Haug}
}

@Article{RW_Model_imp,
AUTHOR = {Segovia, Mariana and Garcia-Alfaro, Joaquin},
TITLE = {Design, Modeling and Implementation of Digital Twins},
JOURNAL = {Sensors},
VOLUME = {22},
YEAR = {2022},
NUMBER = {14},
ARTICLE-NUMBER = {5396},
URL = {https://www.mdpi.com/1424-8220/22/14/5396},
PubMedID = {35891076},
ISSN = {1424-8220},
DOI = {10.3390/s22145396}
}

@article{RW_SoA_survey,
  title={State-of-the-art survey on digital twin implementations},
  author={Liu, YK and Ong, SK and Nee, AYC},
  journal={Advances in Manufacturing},
  volume={10},
  number={1},
  pages={1--23},
  year={2022},
  doi={10.1007/s40436-021-00375-w},
  publisher={Springer}
}

@proceedings{RW_ImpChallenge,
    author = {Trauer, Jakob and Mutschler, Michael and Mörtl, Markus and Zimmermann, Markus},
    title = {Challenges in Implementing Digital Twins – a Survey},
    volume = {Volume 2: 42nd Computers and Information in Engineering Conference (CIE)},
    series = {International Design Engineering Technical Conferences and Computers and Information in Engineering Conference},
    pages = {V002T02A055},
    year = {2022},
    month = {08},
    
    doi = {10.1115/DETC2022-88786},
    url = {https://doi.org/10.1115/DETC2022-88786},
    eprint = {https://asmedigitalcollection.asme.org/IDETC-CIE/proceedings-pdf/IDETC-CIE2022/86212/V002T02A055/6942922/v002t02a055-detc2022-88786.pdf},
}

@Article{RW_challengeApp,
AUTHOR = {Botín-Sanabria, Diego M. and Mihaita, Adriana-Simona and Peimbert-García, Rodrigo E. and Ramírez-Moreno, Mauricio A. and Ramírez-Mendoza, Ricardo A. and Lozoya-Santos, Jorge de J.},
TITLE = {Digital Twin Technology Challenges and Applications: A Comprehensive Review},
JOURNAL = {Remote Sensing},
VOLUME = {14},
YEAR = {2022},
NUMBER = {6},
ARTICLE-NUMBER = {1335},
URL = {https://www.mdpi.com/2072-4292/14/6/1335},
ISSN = {2072-4292},
DOI = {10.3390/rs14061335}
}

@article{RW_survey_DTandCPS,
  title={Digital-twins towards cyber-physical systems: a brief survey},
  author={Parnianifard, Amir and Jearavongtakul, Siwanart and Sasithong, Pruk and Sinpan, Nitinun and Poomrittigul, Suvit and Bajpai, Ambar and Vanichchanunt, Pisit and Wuttisittikulkij, Lunchakorn},
  journal={Engineering Journal},
  volume={26},
  number={9},
  pages={47--61},
  doi={10.4186/ej.2022.26.9.47 },
  year={2022}
}

@ARTICLE{RW_DT_construction,
AUTHOR={Ammar, Ashtarout and Nassereddine, Hala and AbdulBaky, Nadine and AbouKansour, Anwar and Tannoury, Juliano and Urban, Harald and Schranz, Christian},   
TITLE={Digital Twins in the Construction Industry: A Perspective of Practitioners and Building Authority},      
JOURNAL={Frontiers in Built Environment},      
VOLUME={8},           
YEAR={2022},      	  
URL={https://www.frontiersin.org/articles/10.3389/fbuil.2022.834671},       
DOI={10.3389/fbuil.2022.834671},      
ISSN={2297-3362},   
}
\end{document}